\def\papermode{preprint}
\providecommand{\papermode}{submit}
\newif\iftwocol
\def\scpreprintmode{preprint}\edef\scgivenmode{\papermode}
\ifx\scpreprintmode\scgivenmode \twocoltrue \fi
\iftwocol
  \documentclass[final,5p]{elsarticle}
\else
  \documentclass[preprint,12pt]{elsarticle}
\fi

\usepackage[T1]{fontenc}
\usepackage{amsmath}
\iftwocol
  \usepackage{newtxtext,newtxmath}   
\else
  \usepackage{lmodern}
  \usepackage{amssymb}
\fi
\usepackage{microtype}
\usepackage{graphicx}
\usepackage{booktabs}
\usepackage{xcolor}
\usepackage[colorlinks=true,linkcolor=blue!50!black,citecolor=blue!50!black,urlcolor=blue!50!black]{hyperref}

\biboptions{sort&compress}

\newcommand{\oc}{O$-$C}
\newcommand{\dtp}{\ensuremath{\delta t/P}}
\newcommand{\Lsun}{\ensuremath{L_\odot}}
\newcommand{\Msun}{\ensuremath{M_\odot}}
\newcommand{\Rsun}{\ensuremath{R_\odot}}
\newcommand{\starseti}{\textsc{starseti}}
\newcommand{\bcep}{\ensuremath{\beta}~Cephei}
\newcommand{\pbeam}{\ensuremath{P_{\rm beam}}}

\newsavebox{\fitbox}
\newcommand{\fittable}[1]{\sbox{\fitbox}{#1}\ifdim\wd\fitbox>\linewidth\resizebox{\linewidth}{!}{\usebox{\fitbox}}\else\usebox{\fitbox}\fi}

\iftwocol
  \AtBeginDocument{\renewenvironment{highlights}{\setbox0=\vbox\bgroup\begin{itemize}}{\end{itemize}\egroup}}
\fi

\iftwocol
\else
\fi

\journal{Acta Astronautica}

\begin{document}

\begin{frontmatter}

\title{Listening to the Standard Candles: A Survey-Scale Search for Phase-Coded Beacons in Cepheids, RR~Lyrae and $\beta$~Cephei Stars}

\author{James Guillochon}
\ead{guillochon@gmail.com}
\affiliation{Independent Researcher}

\begin{abstract}
Learned et al. (2008) proposed that a civilization could use Cepheids and RR Lyrae stars, the most closely watched clocks in astronomy, as beacons, firing pulsations early with a particle beam to write a message into the timing residuals. The one search to date examined a single Kepler star. We describe Star SETI, an automated pipeline that selects Cepheids, RR Lyrae and $\beta$ Cephei stars from Gaia DR3, measures per-cycle (TESS, Kepler) or block-averaged (ASAS-SN, ZTF) observed-minus-computed timing residuals, tests them for non-white structure, decodes them as a bit stream under the cumulative (``staircase'') modulation of a one-sided trigger, and matches the words to catalogs of recognizable sequences. A resonant antineutrino beam grazing the base of the driving zone needs hundreds to tens of thousands of times less power than a central beam, and driving an RR Lyrae star takes most of its light. Injections into real light curves show that only RR Lyrae stars are detectable near the amplitude that Learned et al. proposed, with half of the codes recovered at 0.2--0.4 per cent of a period per bit in TESS; the ground-based surveys detect only codes held for tens of cycles per bit and decode none. We time 97,725 light curves of 77,596 stars and vet 38,230 timing anomalies; 6823 have timing structure that the vetting cannot attribute to a known cause, all but eight of them on block-averaged timing to which only part of the vetting applies. The highest-ranked are non-radial modes in first-overtone RR Lyrae stars, Blazhko modulation and mode beating in $\beta$ Cephei stars, and no decoded word survives a false-discovery correction. We find no beacon. That bounds only beacons: a perfectly compressed message is indistinguishable from noise, so a civilization communicating efficiently rather than advertising could leave residuals that pass all of our tests, in stars that we cannot single out.
\end{abstract}

\begin{highlights}
\item A one-sided trigger writes a beacon into a pulsator's O$-$C as a cumulative staircase.
\item A resonant antineutrino beam grazing the driving zone cuts the power by $10^{2}$--$10^{4}$.
\item In injections, only TESS RR Lyrae codes are detected at the proposed amplitudes.
\item 97,725 light curves of 77,596 Cepheids, RR Lyrae and $\beta$ Cephei stars are timed.
\item No decoded word survives a false-discovery correction; we find no beacon.
\end{highlights}


\end{frontmatter}

\section{Introduction}
\label{sec:intro}

The search for extraterrestrial intelligence (SETI) began as a search for deliberately transmitted radio signals \citep{Cocconi1959, Drake1961}. For its first half-century it was largely a radio and then an optical enterprise \citep{Tarter2001, Reines2002}. Its vocabulary has since broadened to \emph{technosignatures}, any observable consequence of technology \citep{Wright2018nomenclature, Wright2018taxonomy, Sheikh2020}, which fall into three broad kinds: communicative signals deliberately sent, leakage that escapes as a by-product of a civilization's activities, and artifacts whose presence is detectable regardless of any intent to be seen. Within the communicative category, a \emph{beacon} is a signal designed to be found by a receiver who does not yet know where or how to look, so its defining property is not information capacity but conspicuousness: it must be bright enough, persistent enough and located in a place natural enough that an unprepared observer will stumble on it \citep{Benford2010, Arnold2013, Messerschmitt2015, Wright2021strategies}. That requirement pushes beacon design toward \emph{Schelling points}, choices that both parties can reach independently, and much of the literature on beacons is an argument about which choices those are: magic frequencies from the 21~cm line to Planck-scale multiples \citep{Cocconi1959, Wright2021strategies}, the times and geometry of planetary transits \citep{Arnold2005, Kipping2016, Wright2016megastructures}, or the focal lines of stellar gravitational lenses \citep{Hippke2020network, Kerby2021, Gertz2022}.

A second thread asks not where a signal would be placed but what a civilization with a large energy supply could do to the objects around it. The Dyson sphere \citep{Dyson1960} and the Kardashev scale \citep{Kardashev1964} are the canonical examples, and their infrared and optical signatures were worked out almost immediately \citep{Sagan1966, Shklovskii1966}. Whole-sky searches for that waste heat have since been carried out with IRAS \citep{Carrigan2009}, WISE \citep{Wright2014ghat, Garrett2015, Lacki2019, Chen2021} and Gaia \citep{Zackrisson2018, Suazo2022, Suazo2024}, the scale itself has been re-examined \citep{Cirkovic2015, Wright2020dyson}, and the variants (spheres around white dwarfs, spheres visible outside the infrared, the feedback of a sphere on its host star, and the thermodynamic limits on a sphere used as an engine or computer) have each been treated \citep{Semiz2015, Osmanov2018, Huston2022, Wright2023thermo}. A related family of proposals looks for technology in the light curves and spectra of otherwise ordinary systems: artificial transiting objects and stellar engines \citep{Arnold2005, Forgan2013, Korpela2015, Wright2016megastructures}, industrial pollution, artificial spectral edges and city lights in planetary atmospheres \citep{Lin2014, Lingam2017, Beatty2022, HaqqMisra2022}, and stars that vanish from or appear in archival plates \citep{Villarroel2016}. Between these threads lies the idea of a \emph{stellar beacon}: a civilization that already harvests a substantial fraction of a star's output could also use the star as a transmitter, either by shaping what the star emits \citep{Wright2016megastructures} or by modulating the star itself. Stars are visible across the Galaxy and beyond, they are already cataloged by every astronomer, and a change in their behavior is exactly the kind of anomaly that survey astronomy exists to flag. The cost of such beacons is high, but \citet{Wright2022case} argue that this is the wrong axis on which to dismiss them: technosignatures produced by long-lived, energy-rich societies may be the most detectable and least ambiguous of all, precisely because they are large.

Two episodes from the past decade show what happens when survey astronomy does flag such an anomaly, and both bear on the search described here. The irregular deep dimmings of KIC~8462852 \citep{Boyajian2016} prompted dedicated optical and radio SETI follow-up \citep{Schuetz2016, Lipman2019} before circumstellar dust emerged as the natural explanation; and a claimed detection of periodic spectral modulations in a small fraction of Sun-like stars \citep{Borra2016} was not reproduced in independent high-resolution spectra \citep{Isaacson2019} and was traced to the non-random spacing of absorption lines \citep{Hippke2019lines}. In both cases the anomaly was real as a statistical statement and natural as a physical one, and the work that settled it was a careful accounting of the astrophysical and instrumental ways the statistic could be produced. Section~\ref{sec:hippke} describes the one prior test of the idea pursued here, which followed the same arc, and Section~\ref{sec:beating} describes how the present search is built to avoid repeating it.

\subsection{The Cepheid Galactic Internet}
\label{sec:learned}

\citet{Learned2008} made the stellar-beacon idea concrete. A classical Cepheid is a natural Schelling point: it is a radial pulsator whose period is measured to a part in $10^{6}$ by anyone who uses it as a distance indicator, it is visible to $\sim$20~Mpc with current human technology, and a century of archival timing already exists for the brightest examples. Cepheids pulsate through the $\kappa$-mechanism operating in the partial ionization zone of He~II, a thin layer just below the photosphere, and the onset of each expansion is a threshold event: near the phase at which the star would fire on its own, a modest injection of energy into that layer can trip the cycle early. \citet{Learned2008} proposed that a civilization could ``tickle'' the star at the right moment with a beam of energetic particles. They favored neutrinos, following their own analysis of neutrinos as a galactic communication medium \citep{Learned2009}. Neutrinos of $\sim$TeV energies penetrate the envelope and deposit energy in the interior, and so advance the next pulse by a small fraction of a period. A sequence of such advances, one bit per cycle, is a message written into the star's timing residuals, conventionally called \oc{}. For each pulsation cycle, \oc{} is the \emph{observed} time of a fixed feature of the light curve (a maximum, or the fit of a template) minus the time \emph{computed} from a linear ephemeris $t_0 + nP$, so that a perfectly periodic star has \oc{} identically zero and anything else is structure to be explained. A message written in a recognizable pattern (primes, a counting sequence, a mathematical constant) would reveal its artificial origin to anyone who reads it.

The proposal has been cited in the technosignature literature mostly as an example of the breadth of the beacon class \citep{Wright2018taxonomy, Hippke2018carriers}. \citet{Edmondson2003} had earlier proposed pulsars as natural beacons, and \citet{Chennamangalam2015} developed the transposition, since a pulsar's radio pulses can be modulated far more cheaply than a star can be driven; \citet{Lacki2015} placed it among the things a society commanding Planck-scale energies might attempt; \citet{Hippke2018carriers}, benchmarking information carriers, rated neutrinos a poor medium for point-to-point communication because of the enormous detectors they require, which is a reason a civilization might prefer to use them where the \emph{star} is the detector. More recent beacon proposals encode information in the architecture of planetary systems rather than in stellar timing \citep{Clement2022, Raymond2023}, and the Breakthrough Listen Exotica Catalog \citep{Lacki2021} formalizes the principle that every class of astrophysical object, pulsating stars included, deserves at least one look with technosignatures in mind.

\subsection{The one search to date}
\label{sec:hippke}

\citet{Hippke2015} carried out the only dedicated test of the idea. Working from \emph{Kepler} photometry \citep{Borucki2010}, they measured the cycle-to-cycle period sequence of the first-overtone RR~Lyrae star KIC~5520878 and found that its autocorrelation function was elevated at lags that are prime numbers at a nominal confidence of 99.8 per cent. The result was traced to a natural cause: the star pulsates in two modes with a period ratio near $\sqrt{2}$ \citep{Hippke2014sqrt2}, and the beating of an irrational period ratio through the cycle-to-cycle sequence produces exactly the kind of structured autocorrelation that a naive pattern search picks up. The dynamics turned out to be interesting in their own right, as a first astrophysical example of a strange non-chaotic attractor \citep{Lindner2015}, but as a SETI test it was a null result. Its most durable contribution was procedural: \citet{Hippke2015} recommended that the search be automated and run as a by-product of routine time-series analysis over every archival and future photometric survey.

That recommendation was made in 2015 with a single space telescope and a handful of RR~Lyrae stars in hand. This paper describes an attempt to carry it out across the Galactic population of the three classes of pulsator for which it is best suited, using the survey data that have accumulated since. Section~\ref{sec:physics} revisits the physics of the proposed transmitter and derives the observable that a search should look for. Section~\ref{sec:surveys} summarizes what a decade of space and ground-based surveys has established about the intrinsic timing behavior of Cepheids, RR~Lyrae and \bcep{} stars, which is the astrophysical foreground for any such search. Section~\ref{sec:method} describes the pipeline, its statistics and its measured sensitivity. Section~\ref{sec:viewer} describes the web interface that publishes the results, and Section~\ref{sec:candidates} presents the highest-ranked stars. Section~\ref{sec:compression} sets out what a null result of this kind cannot exclude, and Section~\ref{sec:conclusions} gives our conclusions.

\section{The Physics of a Driven Pulsator}
\label{sec:physics}

\citet{Learned2008} sketched the mechanism and its energy budget but, as they note, stopped short of stellar modeling. This section works through what a transmitter has to do, because the answer determines both the waveform that a search should look for and the population in which a search has the best chance. Throughout, the reference stars are a classical Cepheid ($M=5\,\Msun$, $R=50\,\Rsun$, $L=3000\,\Lsun$, $P=10$~d), a fundamental-mode RR~Lyrae star ($0.7\,\Msun$, $5\,\Rsun$, $50\,\Lsun$, $0.57$~d) and a \bcep{} star ($10\,\Msun$, $7\,\Rsun$, $15000\,\Lsun$, $0.20$~d). The numbers are collected in Table~\ref{tab:physics}.

\subsection{One-sided triggering}
\label{sec:onesided}

The $\kappa$-mechanism drives pulsation because the opacity of the partial-ionization zone rises on compression, so the layer absorbs heat when compressed and releases it on expansion, doing net work on the oscillation each cycle \citep{Buchler1997, Gastine2008}. Near the end of contraction, the resulting nonlinear limit cycle passes through a phase at which the driving layer is primed and sensitive to additional heat: energy delivered then trips the expansion early. The same energy delivered at other phases is simply absorbed into the thermal pool of the envelope and radiated away. In this picture there is no corresponding way to make the star fire \emph{late}. A phase code written this way is therefore \emph{one-sided}: a set bit advances the clock, a clear bit leaves it alone. This is the model \citet{Learned2008} describe, and we adopt it for our injections and decoders. It is an assumption rather than a result: the phase response of a nonlinear limit-cycle oscillator to a brief perturbation is in general biphasic, and no phase-response curve of a $\kappa$-driven pulsator has been computed. We therefore do not rely on the sign. The matched filter of Section~\ref{sec:decode} tests both polarities, and the extra trial is included in its $p$-value; the fold decoder does not depend on polarity.

\subsection{The O-C staircase}
\label{sec:staircase}

A pulsating star is an autonomous nonlinear oscillator on a limit cycle, and on a limit cycle the phase is the marginal direction: perturbations in amplitude decay back onto the cycle, but a perturbation in phase has zero Lyapunov exponent and is never undone \citep{Buchler1997}. A cycle triggered early is not followed by a compensating late one; the next unperturbed pulse arrives one period after the \emph{early} one, and every subsequent pulse inherits the advance. The step itself is not instantaneous. The heat that a kick deposits relaxes on the thermal time of the driving zone, which is of order the period (that is why the zone drives at all). The cycles after a kick therefore run with a slightly altered period, and the advance accrues over one or a few cycles rather than arriving all at once. The clean step assumed below is an idealization; the injection stage of Section~\ref{sec:injection} can give each kick an exponential tail, and the cost of a one-cycle tail is measured there. This is the behavior that \citet{Eddington1929} inferred for long-period variables and that has since been documented for Cepheids and RR~Lyrae stars: pulsator \oc{} diagrams behave as random walks assembled from accumulated period fluctuations, not as bounded jitter about a fixed ephemeris \citep{Turner2006, Csornyei2022}.

The consequence for a beacon is that the \oc{} does not trace the bit sequence itself. If $b_k\in\{0,1\}$ is the bit of cycle $k$ and $\delta t$ the advance per set bit, the observed timing residual after cycle $n$ is the running sum
\begin{equation}
  (\mathrm{O}-\mathrm{C})_n = -\,\delta t \sum_{k\le n} b_k ,
  \label{eq:staircase}
\end{equation}
a one-sided staircase whose \emph{steps}, the first difference of the \oc{}, are the bits. Three things follow. First, a constant duty cycle $d$ produces a linear ramp of slope $-d\,\delta t$ per cycle, which any period search absorbs into a slightly shorter period; the detrended \oc{} then shows steps of $+d\,\delta t$ on clear bits and $-(1-d)\,\delta t$ on set bits. Second, a decoder must threshold increments rather than levels; a level-thresholding decoder applied to a staircase returns the integral of the message. Third, each bit is now encoded in the difference of two noisy timings, and a step across a gap in the data is lost, so the physical waveform is harder to detect than the square wave that a naive search would inject. Our injection tests (Section~\ref{sec:injection}) quantify the penalty at a factor of two to four in amplitude.

\subsection{Envelope heating}
\label{sec:heating}

Energy deposited in the envelope expands it and, since $P\propto\bar\rho^{-1/2}$, \emph{lengthens} the period, opposite in sign to the advance being signalled. A Cepheid driven at $\dtp=10^{-3}$ per cycle at unit gain and a duty cycle of one half drifts at $+0.25$~s~yr$^{-1}$, far inside the $1$--$100$~s~yr$^{-1}$ range of observed evolutionary period changes \citep{Turner2006, Neilson2015, Csornyei2022}, and after a Kelvin--Helmholtz time, $\tau_{\rm KH}=G_{\rm N}M^2/RL$ with $G_{\rm N}$ the gravitational constant (about $5\times10^{3}$~yr for the reference Cepheid and $6\times10^{4}$~yr for the RRab star), the star settles at a new, slightly larger equilibrium radius. For the search the back-reaction matters in one way only: a constant $\dot P$ is exactly the quadratic \oc{} term that our detrending removes, so it can neither hide a code nor imitate one.

\subsection{Energy per bit}
\label{sec:cost}

Write the trigger energy as
\begin{equation}
  E_{\rm bit} = \frac{L\,\delta t}{G\, f_{\rm abs}\, f_{\rm zone}} ,
  \label{eq:ebit}
\end{equation}
where $L\,\delta t$ is the stellar energy the phase advance is worth, $G$ is the gain that \citet{Learned2008} invoke when they call the star ``a huge signal amplifier'' and decline to quantify, $f_{\rm abs}$ is the fraction of the beam the star absorbs at all, and $f_{\rm zone}$ is the fraction of the absorbed energy that is deposited where it can trigger the pulse. We take $G=1$ throughout as a reference value; every figure below scales as $1/G$. It is not a conservative choice. For an autonomous oscillator an impulsive energy injection $\Delta E$ at the sensitive phase shifts the phase by $\Delta\phi/2\pi\approx Z\,\Delta E/E_{\rm puls}$, with $E_{\rm puls}$ the energy in the pulsation and $Z$ a phase sensitivity of order unity at the best phase, so $G\approx Z\,LP/E_{\rm puls}$: the gain is the ratio of the energy that the star radiates per cycle ($LP$, $10^{43}$~erg for the reference Cepheid and $10^{40}$~erg for the RRab) to the energy stored in its pulsation. In the limit cycle the driving zone's work per cycle balances the dissipation, $W=2\kappa E_{\rm puls}$ with $\kappa$ the growth rate per cycle, of order $10^{-2}$ for the radial modes of RR~Lyrae stars in hydrodynamic models \citep{Bono1997, Kollath2011}, and $W$ is a fraction $\epsilon$ of the energy that the star radiates per cycle, a few per cent to a few tenths, so $G\approx2Z\kappa/\epsilon$, between 0.1 and 1 at $Z=1$. A second route agrees. The driving zone sits where the thermal time of the layers above it is of order the period (Section~\ref{sec:staircase}), so those layers hold a thermal energy of order $LP$; at the surface velocity amplitude of an RRab star, $\sim$40~km~s$^{-1}$, their kinetic energy is comparable to their thermal energy, so $E_{\rm puls}\gtrsim LP$ and $G\lesssim Z$ from this side too. Only the relaxation-oscillator picture, in which the zone fires when its heat content crosses a threshold so that $\Delta E=L\,\delta t$ advances it by $\delta t$, gives $G\approx1$ or a little more. Until the phase-response curve noted in Section~\ref{sec:onesided} has been computed, $G=1$ should be read as the upper end of a range that extends an order of magnitude below it.

\begin{figure*}[tp]
  \centering
  \includegraphics[width=\textwidth]{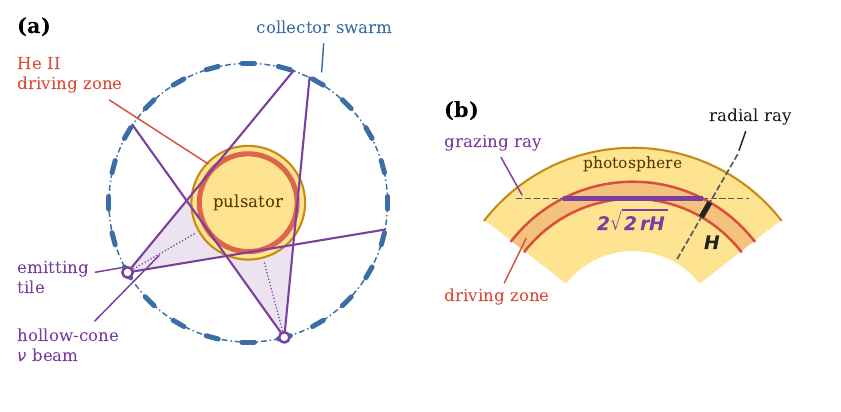}
  \caption{Schematic of the transmitter geometry (not to scale). (a) The host pulsator with its He~II driving zone (red shell), at radius $r_{\rm zone}$ and of thickness $\sim H$, just below the photosphere; its depth is exaggerated, since $H/R$ of a few per cent would place it at the edge of the disk. A partial Dyson swarm of collectors (blue) intercepts a fraction $f_{\rm coll}$ of the star's sky at a stand-off of $\sim40\,R$, drawn much closer here. It collects the starlight $L$ and converts it to a beamed neutrino flux at efficiency $\eta$. One emitting tile (open circle) fires a hollow-cone neutrino beam whose edges graze the base of the driving zone at impact parameter $b=r_{\rm zone}$. Each ray samples only radii $\ge r_{\rm zone}$, so the beam crosses the zone and the tenuous layers above it, leaves on the far side, and never illuminates the core; the axis of the cone (dotted) is empty. Emitters distributed over $4\pi$ each illuminate their own limb, and the sum is a spherical shell. (b) Advantage of the geometry. A radial ray through the star crosses a shell of thickness $H$ twice, a path of $2H$. A ray grazing the shell at radius $r$ crosses it along the chord $2\sqrt{2rH}$, drawn in full, an enhancement of $\sqrt{2R/H}\approx7$ for a Cepheid, $\approx8$ for an RRab star and $\approx14$ for a \bcep{} star. That enhancement is what keeps the required beam power tolerable once the core is no longer the absorber. Produced by \texttt{figures/make\_geometry.py}.}
  \label{fig:geometry}
\end{figure*}

\paragraph{Transparency, $f_{\rm abs}$.} For a beam through the disk, $f_{\rm abs}=\Sigma/\Lambda$ with $\Sigma=M/\pi R^2$ the disk-averaged column and $\Lambda=m_p/\sigma_{\nu N}(E)$ the interaction column. At 1~TeV, $\sigma_{\nu N}\simeq7\times10^{-36}$~cm$^2$ gives $\Lambda=2.4\times10^{11}$~g~cm$^{-2}$, consistent with the ``$10^6$~km at solar density'' quoted by \citet{Learned2008}. Since $\Sigma\propto M/R^2$, the absorbed fraction at fixed energy scales with surface gravity: $f_{\rm abs}\approx10^{-3}$ for a Cepheid, $1.5\times10^{-2}$ for an RRab, $0.1$ for a \bcep{} star. Everything in the classical instability strip is transparent at 1~TeV. The cross-section does not rise without limit: the deep-inelastic cross-section is linear in $E$ only to a few TeV, and above $\sim$10~PeV the $W$ propagator turns it over to {\relpenalty=200 $\sigma_{\nu N}\simeq7.8\times10^{-36}\,(E/{\rm GeV})^{0.363}$}~cm$^2$ \citep{Gandhi1996, Gandhi1998}, the sublinear regime that IceCube has since measured between 18~TeV and 10~PeV \citep{Bustamante2019, IceCube2021xsec}. With that law, the energy at which the disk is one interaction length deep, $\tau=1$, is $\sim$100~PeV for a Cepheid, 70~TeV for an RRab and 9~TeV for a \bcep{} star. All three lie inside the spectrum nature produces: IceCube's astrophysical neutrinos reach a few PeV \citep{IceCube2013pev, IceCube2014} and KM3NeT has detected one at $\sim$220~PeV \citep{KM3NeT2025}. \citet{Learned2008} chose 1~TeV to argue penetration to the core; for signalling one wants $\tau\approx1$, at which $f_{\rm abs}\approx0.6$ by construction.

\paragraph{Deposition, $f_{\rm zone}$.} Neutrinos deposit energy in proportion to the mass they traverse, so a beam through the center of the star puts nearly all of its energy in the core and almost none in the driving zone. How little depends on the envelope structure, and the driving zone is far shallower than a hydrostatic estimate suggests. We define the target layer as the \emph{base of the driving zone}, the depth at which the thermal time of the overlying layers equals one period \citep[the transition-region criterion of][]{Cox1980}: above it, heat leaks out within a cycle and does not drive; below it, heat is stored for many cycles and reaches the zone as a slow luminosity excess, so this is the deepest layer at which a kick produces a step rather than a ramp (Section~\ref{sec:staircase}). For the RRab we read the structure off a MESA horizontal-branch model of $0.63\,\Msun$, $46\,\Lsun$ and 6725~K \citep{Zenodo15741530}: the base of the He~II zone sits at 55{,}000~K, 0.91 of the radius, with a column of 750~g~cm$^{-2}$ above it, a density of $6\times10^{-8}$~g~cm$^{-3}$ and a scale height $H/R=0.03$, so the mass above it is $f_{\rm above}\approx8\times10^{-7}$ of the star. A grey radiative envelope integrated inward from the photosphere reproduces that model to a factor of two and gives $f_{\rm above}\approx10^{-4}$ for the Cepheid, whose base sits at 58{,}000~K, and $1.4\times10^{-6}$ for the \bcep{} star, whose driving zone is the iron-opacity bump near $2\times10^{5}$~K (Table~\ref{tab:physics}). The same criterion is what makes the gain estimate above self-consistent, since the layers above the base hold a thermal energy of order $LP$ by construction. For a central beam $f_{\rm zone}=f_{\rm above}$, and the beam power required to sustain a code of duty cycle $d$ is
\begin{equation}
  \frac{\pbeam}{L} = \frac{d\,(\dtp)}{G\, f_{\rm abs}\, f_{\rm zone}} ,
  \label{eq:pbeam}
\end{equation}
which at $\dtp=10^{-3}$, $d=0.5$ and $\tau=1$ comes out at $\pbeam/L\approx8$ for the Cepheid, 600 for the \bcep{} star and 1000 for the RRab. A central-beam transmitter at unit gain runs at ten to a thousand times the luminosity of the star that it drives, which no collector around that star can supply. The next subsection describes a geometry that avoids this problem.

\subsection{An annular beam at the limb}
\label{sec:annulus}

The fix is geometric (Figure~\ref{fig:geometry}). A ray at impact parameter $b$ samples only radii $\ge b$. A hollow cone of rays with $b$ at the base of the driving zone therefore passes through the zone and the layers above it and nothing else. The cost is a smaller column. A grazing ray crosses a shell of thickness $H$ along a chord $2\sqrt{2rH}$, whereas a radial ray through the star meets the shell twice and crosses only $2H$. The enhancement, $\sqrt{2R/H}$, is about 7 for a Cepheid ($H/R\approx0.04$ at the base of the zone), 8 for an RRab and 14 for a \bcep{} star ($H/R\approx0.01$). Even so, the column through the envelope alone is far smaller than the column through the disk, by {\relpenalty=200 $\Sigma_{\rm disk}/\Sigma_{\rm graze}=2/(\sqrt{2R/H}\,f_{\rm above})$}, about $3\times10^{3}$, $3\times10^{5}$ and $10^{5}$ ($9\times10^{4}$, $1.2\times10^{4}$ and $2.5\times10^{5}$~g~cm$^{-2}$ against $2.6\times10^{8}$, $3.7\times10^{9}$ and $2.7\times10^{10}$). No grazing path reaches $\tau=1$ at any energy a neutrino can have. At 1~EeV, where $\sigma_{\nu N}\simeq1.4\times10^{-32}$~cm$^2$, the three limbs absorb $8\times10^{-4}$, $10^{-4}$ and $2\times10^{-3}$ of the beam. The resonant channel is the one that works. A monochromatic $\bar\nu_e$ beam at the Glashow resonance, 6.3~PeV, sees $\sigma_{\bar\nu_e e}\simeq5\times10^{-31}$~cm$^2$ per electron, the largest cross-section a neutrino has, or $\Lambda\approx4\times10^{6}$~g~cm$^{-2}$ in a hydrogen--helium envelope \citep{Gandhi1998}; IceCube has detected one such event \citep{IceCube2021glashow}, and a beam needs an energy spread of only a few per cent to sit on the resonance. On the grazing columns it gives $\tau=f_{\rm abs}=0.023$ for the Cepheid, $3\times10^{-3}$ for the RRab and $0.064$ for the \bcep{} star, thirty times what the same limbs absorb from the deep-inelastic beam at 1~EeV. Because every limb is optically thin, the deposition follows the local density along the chord, and 84 per cent of it lands within one scale height above the tangent point, inside the zone whose base the ray grazes: $f_{\rm zone}\approx1$, and the deposition penalty of $1/f_{\rm above}$, between $10^{4}$ and $10^{6}$, is cancelled outright. The beam power at unit gain (Equation~\ref{eq:pbeam}) is then $\pbeam/L=d\,(\dtp)/f_{\rm abs}$: at $\dtp=10^{-3}$ and $d=0.5$, $0.022$ for the Cepheid, $0.16$ for the RRab and $8\times10^{-3}$ for the \bcep{} star, hundreds to tens of thousands of times less than the central beam. The \bcep{} stars stand out because their thin, high-gravity envelopes give both the largest chord enhancement and the densest limb; the RR~Lyrae stars, with the most tenuous limb of the three, are the class the geometry helps least.

Emitters distributed over the sphere each illuminate their own limb, so the sum over the swarm is a spherical shell; a sparse array would also drive $\ell\ge2$ modes, and excess non-radial power phase-locked to the modulation would be its observational signature.

\subsection{Limits on energetics}
\label{sec:swarm}

At $\pbeam/L$ of $10^{-2}$ to $10^{-1}$ the only plausible power source is the star itself, so the transmitter is a collector array in orbit at a stand-off of tens of stellar radii, where aiming the cone at the limb needs arcminute rather than micro-arcsecond accuracy.

A collector intercepting a fraction $f_{\rm coll}$ of the star's sky and converting starlight to a beamed neutrino flux at efficiency $\eta$ supplies $\pbeam=\eta f_{\rm coll}L$, so
\begin{equation}
  \begin{aligned}
    f_{\rm coll} &= \frac{\pbeam}{\eta L} \\
    \Longrightarrow\quad \frac{\delta t}{P}
      &\le \frac{\eta\, f_{\rm coll}}{d}\, G f_{\rm abs} f_{\rm zone} .
  \end{aligned}
  \label{eq:ceiling}
\end{equation}
We let $\eta$ approach unity rather than fix it at a tenth. The question is whether a transmitter powered by its own star could produce a signal that the data could detect, and that question is best posed at the physical limit. The sphere fractions at $\eta=1$ are those of Table~\ref{tab:physics}. Because of the annular geometry, the unquantified gain $G$ is critical only in the central-beam picture. Equation~\ref{eq:ceiling} caps the code amplitude a self-powered transmitter can produce. We take the transmitter to have no energy store, so that it can put into one kick only what it collected during the preceding cycle, $\eta f_{\rm coll}LP$, and its advance per set bit obeys $\dtp\le\eta f_{\rm coll}\,Gf_{\rm abs}f_{\rm zone}$ whatever the duty cycle (the $d=1$ case of Equation~\ref{eq:ceiling}). With a complete sphere at unit efficiency that is $\dtp\le f_{\rm abs}$: $0.023$ for a Cepheid, $3\times10^{-3}$ for an RRab and $0.064$ for a \bcep{} star. We refer to that number below as the \emph{$G=1$ kick ceiling}; it is the largest code that a storage-free transmitter at unit gain can send using all of its star's light. It scales with $G$ (Section~\ref{sec:cost}), and it is the scale against which the sensitivity of a search should be judged. A transmitter that covers a twentieth of the sky at ten per cent efficiency falls two hundred times below it.

\begin{table*}[tp]
\centering
\caption{Reference stars and the cost of driving them at unit gain \label{tab:physics}}
\scriptsize
\setlength{\tabcolsep}{2.6pt}
\fittable{\begin{tabular}{lccccccccccc}
\toprule
 & & \multicolumn{5}{c}{Base of the driving zone} & \multicolumn{2}{c}{Central beam} & \multicolumn{3}{c}{Resonant annulus} \\
\cmidrule(lr){3-7} \cmidrule(lr){8-9} \cmidrule(lr){10-12}
Class & $\log g$ & $T$ (K) & $\Sigma_{\rm above}$ & $f_{\rm above}$ & $H/R$ & $\Sigma_{\rm graze}$ & $E_{\tau=1}$ & $\pbeam/L$ & $f_{\rm abs}$ & $\pbeam/L$ & $f_{\rm coll}$ ($\eta=1$) \\
\midrule
Cepheid  & 1.74 & 58{,}000  & $6.5\times10^{3}$ & $1.0\times10^{-4}$ & 0.042 & $9.0\times10^{4}$ & 100~PeV & 8.3  & 0.023 & 0.022 & 2.2\% \\
RRab     & 2.89 & 55{,}000  & $7.3\times10^{2}$ & $8\times10^{-7}$   & 0.030 & $1.2\times10^{4}$ & 70~TeV  & 1040 & $3.0\times10^{-3}$ & 0.16 & 16\% \\
\bcep{}  & 3.75 & 220{,}000 & $9.3\times10^{3}$ & $1.4\times10^{-6}$ & 0.011 & $2.5\times10^{5}$ & 9~TeV   & 600  & 0.064 & $7.8\times10^{-3}$ & 0.8\% \\
\bottomrule
\end{tabular}}
\begin{minipage}{\textwidth}
\vspace{1pt}
\small \textsc{Note}---The base of the driving zone is the layer whose overlying thermal time equals one period; $\Sigma_{\rm above}$ (g~cm$^{-2}$) is the column above it, $f_{\rm above}$ the mass fraction above it, and $\Sigma_{\rm graze}=\Sigma_{\rm disk}f_{\rm above}\sqrt{2R/H}/2$ the column along a chord grazing it. The RRab row is read off a MESA horizontal-branch model \citep{Zenodo15741530}; the other two come from a grey radiative envelope that reproduces that model to a factor of two. Beam powers are for $\dtp=10^{-3}$ per set bit at a duty cycle of $0.5$ and $G=1$, with $f_{\rm zone}=f_{\rm above}$ for the central beam at its $\tau=1$ energy (cross-section of \citealt{Gandhi1998}) and $f_{\rm zone}=1$ for the resonant annulus at the Glashow resonance, whose absorbed fraction is the optical depth of the grazing column at $\Lambda=4\times10^{6}$~g~cm$^{-2}$. The ceiling is the largest $\dtp$ a storage-free transmitter with a complete sphere at unit efficiency can send, $f_{\rm abs}$ (Section~\ref{sec:swarm}). The derivation and the model comparison are in the \texttt{research/} notes released with the code.
\end{minipage}
\end{table*}

\subsection{Costs by stellar class}
\label{sec:whichstars}

Nothing above argues against Cepheids as the \emph{target} of a beacon meant to be seen from far away: they have sixty times the luminosity of an RRab, and $\dtp=10^{-3}$ is an 860~s shift rather than a 50~s one. But a cost-conscious transmitter would pick RR~Lyrae stars, which are 140 times cheaper per bit than a Cepheid ($3\times10^{39}$ against $4\times10^{41}$~erg at $\dtp=10^{-3}$) and five times cheaper than a \bcep{} star, and offer twenty times the bit rate, although against the star's own light the order reverses (Table~\ref{tab:physics}). Section~\ref{sec:injection} shows that the RR~Lyrae stars are also the class to which our data are most sensitive, which is why the null result that follows is mainly a result about RR~Lyrae stars.

\section{Pulsating Stars in the Survey Era}
\label{sec:surveys}

Any search for artificial structure in pulsation timing is a search against the intrinsic timing behavior of the stars, and the decade since \citet{Hippke2015} has produced a far richer picture of that behavior than existed when \citet{Learned2008} wrote. We summarize the parts that bear on the search: how many stars there are and how well they are cataloged, how noisy their clocks are cycle to cycle, and what coherent structure remains once the primary period is removed.

\subsection{Catalogs}
\label{sec:catalogs}

Gaia~DR3 \citep{GaiaDR3} is the first homogeneous all-sky census of the classes we target. Its variability processing \citep{Eyer2023} identified 15{,}021 Cepheids of all types \citep{Ripepi2023} and $\sim$271{,}000 RR~Lyrae stars \citep{Clementini2023}, each with a period, a Fourier decomposition and a mode classification, and its supervised classifier \citep{Rimoldini2023} assigns \bcep{} candidates among 12.4 million variable candidates. About a third of the Cepheids and the majority of the RR~Lyrae stars are Galactic. The classifications have been validated against \emph{Kepler} and K2 photometry \citep{Molnar2018gaia} and are the natural starting point for a target list; they replace the heterogeneous compilations that an earlier search would have had to combine (for \bcep{} stars, the closest analogue is the catalog of \citealt{Stankov2005}). Ground-based survey catalogs of variables from ASAS-SN \citep{Jayasinghe2018, Jayasinghe2019, Christy2023} and ZTF \citep{Chen2020} overlap these samples and add their own light curves.

A census is not a set of clocks. Gaia's epoch photometry \citep{Eyer2023} is what the scanning law delivers, a few tens of transits per source scattered over the 34 months of DR3: enough for a period and a Fourier decomposition, but not for an \oc{}, which needs either an individually resolved maximum or a block of epochs dense enough that a template fit localizes the phase. The timing therefore comes from TESS for per-cycle work and from ASAS-SN and ZTF for block-averaged phases, each with its own bright and faint limits. Of the $\sim$271{,}000 DR3 RR~Lyrae stars, mostly faint halo and bulge objects, only about 8 per cent (21{,}110) pass the $G\le15.5$ cut and carry a TESS identifier; ZTF, whose faint window extends to $G=18.5$, reaches the most (61{,}975), and ASAS-SN reaches 15{,}942 (Table~\ref{tab:targets}). The catalog is all-sky; the search is not.

Gaia~DR4, built from 66 months of data rather than 34 \citep{GaiaDR4docs}, is expected to release epoch photometry far more widely. Its sampling will still be too sparse for per-cycle timing, but twice the DR3 number of transits over 5.5~yr is the regime in which the block-averaged phases of Section~\ref{sec:sparse} work, and stitched to the ASAS-SN and ZTF baselines it would lengthen the coherent baseline over which a slow cumulative code accumulates phase.

\subsection{Intrinsic timing noise}
\label{sec:jitter}

Space photometry of Cepheids has shown that even the ``cleanest'' pulsators have unstable periods on short timescales. The \emph{Kepler} Cepheid V1154~Cyg exhibits period fluctuations of $\sim$20 minutes cycle to cycle with random-walk character \citep{Derekas2012}, later attributed to granulation noise in its extended convective envelope \citep{Neilson2014, Derekas2017}. On long baselines, the period changes of Cepheids are dominated by evolution but carry a stochastic component that the Eddington--Plakidis test isolates \citep{Turner2006, Csornyei2022}.

RR~Lyrae stars are noisier still, in a more structured way. The Blazhko effect modulates amplitude \emph{and} phase over tens to hundreds of days in roughly half of RRab stars \citep{Kolenberg2011, Skarka2020}, with period doubling and additional modes tied to it \citep{Szabo2010, Benko2010, Moskalik2015}; the prototype RR~Lyr shows its pulsation period wandering between 0.5655 and 0.5685~d across a Blazhko cycle \citep{Stellingwerf2013}. The non-Blazhko stars are not quiet either: \citet{Benko2019} found cycle-to-cycle light-curve variations and additional modes in the \emph{Kepler} non-Blazhko sample, and \citet{Benko2025, Benko2026} have recently connected those light-curve variations directly to \oc{} variations, showing that the timing residuals of RR~Lyrae stars carry the imprint of their shape changes, exactly the coupling that a phase-modulation search must control for. Long-term cyclic \oc{} variations that resemble light-travel-time orbits are common and, as the case of Z~CVn shows, can be period instability rather than a companion \citep{Skarka2018}. TESS has now extended these results to a large sample of bright Galactic RR~Lyrae stars with its own mode-content census \citep{Molnar2022}, and K2 has done the same for first-overtone pulsators \citep{Netzel2023}.

\bcep{} stars are a different regime: hot, high-gravity, low-amplitude and almost always multi-periodic. The ASAS-3 study of \citet{Pigulski2008} found long-term period and amplitude changes in a substantial fraction; TESS has multiplied the known sample and characterized their frequency content \citep{Shi2024, Eze2024, Shi2025}, and their period changes are consistent with main-sequence evolution \citep{Neilson2015}. For a timing search their multi-periodicity is the dominant nuisance: the beating of independent modes reads as a large, coherent \oc{} signal unless the modes are removed first.

\subsection{Consequences for a search}
\label{sec:foreground}

Three lessons follow. First, the noise floor of the search is not the per-cycle timing precision but the star's own timing wander. Our TESS injection hosts (Section~\ref{sec:injection}) make this concrete: for a typical TESS Cepheid the \oc{} scatter is 25 times the formal per-cycle timing error, for a \bcep{} star six times, and for an RR~Lyrae star only about twice. RR~Lyrae stars have the clocks closest to white noise, which is a major reason that the search is most sensitive to them. Second, a white-noise null is the wrong null for a red-noise clock. Any statistic calibrated against random permutations of the \oc{} will flag a Cepheid that is merely behaving like a Cepheid, and the vetting has to reject red noise explicitly, with a block bootstrap that preserves the short-range correlation \citep{Kunsch1989}. Third, the coherent residuals that remain once the primary period is subtracted (Blazhko modulation, light-travel-time sinusoids, beating between modes, and the survey's own sampling and systematics periods) are smooth. A designed code is not smooth: its steps come in two sizes and nothing in between. The search is built around that contrast: a designed code has a two-level distribution of increments, whereas every known astrophysical mechanism produces a continuous one.

\section{Method}
\label{sec:method}

The pipeline, \starseti{}, runs in stages. Stage~0 builds the target list from Gaia~DR3 (Section~\ref{sec:targets}). The later stages are light-curve retrieval and per-star timing (Sections~\ref{sec:lightcurves}--\ref{sec:sparse}), whiteness and decodability statistics with a triage decision (Sections~\ref{sec:whiteness}--\ref{sec:triage}), a vetting battery (Section~\ref{sec:vetting}), a ranking score (Section~\ref{sec:score}) and an injection-recovery measurement of the search's completeness (Section~\ref{sec:injection}). Every stage reads from and writes to a single SQLite store, which is also what the public web interface (Section~\ref{sec:viewer}) serves. The source code, the research notes in which the physics and statistics were derived, and the scripts that produced every figure in this paper are released with it. The pipeline was run twice: a test pass and a production pass. Every count in this paper is from the production pass unless the text says so.

\subsection{Targets}
\label{sec:targets}

Targets are drawn from the Gaia~DR3 Specific Object Study tables of Cepheids \citep{Ripepi2023} and RR~Lyrae stars \citep{Clementini2023}, keeping the classical, Type~II and anomalous Cepheids and the RRab, RRc and RRd stars, and from the DR3 supervised classification \citep{Rimoldini2023} for \bcep{} candidates, for which no SOS table exists. Stars in the Magellanic Clouds, M31 and M33 are removed by position, and a magnitude cut of $G\le15.5$ is applied for TESS; ZTF and ASAS-SN use their own windows below. Each survey then receives the subset of targets that it can observe. The TESS Input Catalog \citep{Stassun2019} is cross-matched to obtain TIC identifiers. ASAS-SN Sky Patrol identifiers \citep{Kochanek2017, Hart2023} are found through the survey's own Gaia~DR2 cross-identifications and a $3''$ cone search, restricted to $10.5\le G\le15.5$, between saturation and the useful faint limit. ZTF object identifiers \citep{Bellm2019, Masci2019} are found by a $1.5''$ cone match within the survey's own window, $13\le G\le18.5$, whose bright end is set by saturation (which is why only eight \bcep{} stars have ZTF rows). A ZTF object is one field, quadrant and filter, so a star typically has several identifiers. The Gaia period and peak-to-peak amplitude, where available, are carried on each target row. Table~\ref{tab:targets} gives the resulting counts.

\begin{table}[tp]
\centering
\caption{Target lists and processing status \label{tab:targets}}
\begin{tabular}{lrrr}
\toprule
 & Cepheids & RR Lyrae & \bcep{} \\
\midrule
\multicolumn{4}{l}{\emph{Gaia DR3 targets with a survey identifier}} \\
TESS    & 3242 & 21110 & 1474 \\
ASAS-SN & 1915 & 15942 & 601 \\
ZTF     & 1643 & 61975 & 8 \\
Kepler  & 2    & 23    & 1 \\
\addlinespace
\multicolumn{4}{l}{\emph{Timed light curves (stage 1 completed)}} \\
TESS    & 2759 & 12892 & 1454 \\
ASAS-SN & 1832 & 15128 & 575 \\
ZTF     & 1637 & 61414 & 8 \\
Kepler  & 2    & 23    & 1 \\
\addlinespace
\multicolumn{4}{l}{\emph{Flagged for vetting}} \\
TESS    & 183 & 2684  & 833 \\
ASAS-SN & 251 & 7034  & 68 \\
ZTF     & 400 & 26756 & 0 \\
Kepler  & 1   & 20    & 0 \\
\bottomrule
\end{tabular}

\begin{minipage}{\columnwidth}
\vspace{1pt}
\small \textsc{Note}---Counts from the store on 2026 September 26, excluding the three synthetic controls: 97{,}725 timed light curves of 77{,}596 distinct Gaia sources, 38{,}230 of them flagged. Of the TESS RR~Lyrae rows, 4599 are two-minute-cadence light curves timed per cycle and 8293 are full-frame-image rows timed in blocks (Section~\ref{sec:sparse}); for the \bcep{} stars the split is 1344 and 110. The store's dense count includes one synthetic control per class, which this table excludes. The Kepler rows are named targets only (Section~\ref{sec:beating}); the Kepler field RR~Lyrae sample as a whole is not included and is the obvious dense-mode extension. The 8218 TESS RR~Lyrae stars without a timed light curve have no archived product we could read. Full-frame-image cutouts for them are left to future work.
\end{minipage}
\end{table}

\subsection{Light curves}
\label{sec:lightcurves}

TESS \citep{Ricker2015} light curves are retrieved from MAST with \texttt{lightkurve} \citep{Lightkurve2018}, preferring two-minute (or, for \bcep{} stars, 20~s) cadence products and falling back to full-frame-image pipelines where those are all that exist: the SPOC FFI target lists \citep{Caldwell2020}, the Quick-Look Pipeline \citep{Kunimoto2022}, and the image-subtraction light curves of the T16 project \citep{Hartman2025}. Because \texttt{lightkurve} cannot read the T16 products, we parse them ourselves, using the undetrended magnitudes so that a high-pass filter does not remove the Cepheid pulsation. Quality-flagged cadences are removed and each sector is normalized separately. ASAS-SN \citep{Shappee2014} epochs come from the Sky Patrol~V2 service \citep{Hart2023} in $g$ (the $V$ and $g$ eras are different epochs of the survey and are not mixed), and ZTF epochs from the IRSA light-curve service in $g$ and $r$ together. Raw rows are cached before any cut. The quality mask, band selection and outlier clipping are applied when the cache is read, so one download serves every reprocessing.

Ground-based surveys observe with several cameras or, for ZTF, several fields and filters. Each contributes its own zero point and its own amplitude scale, because a pulsator's amplitude in $r$ is $\sim$0.75 of that in $g$ and blends differ between fields. Both are fitted per camera alongside the template (Section~\ref{sec:template}). Without the scale correction, merging two cameras with no injected jitter produces a spurious jitter detection at four times the gate; with it the statistic returns to zero and the recovered scale ratio is correct to a per cent.

\subsection{Period and template}
\label{sec:template}

For each star the pulsation period is found with a fast Lomb--Scargle periodogram \citep{Press1989} over the class's period window (0.05--60~d for Cepheids, 0.1--2~d for RR~Lyrae, 0.03--1~d for \bcep{}), or within $\pm2$ per cent of the Gaia period where one exists, and refined by a local exact search. The Cepheid window is deliberately wide at the short end: the Gaia period takes precedence wherever it exists, and a $\delta$~Scuti star misclassified as an anomalous Cepheid is timed and vetted like any other rather than dropped. A Fourier template
\begin{equation}
  \begin{aligned}
    F(\phi) &= c_0 + \sum_{k=1}^{n_h}\left[a_k\cos 2\pi k\phi + b_k\sin 2\pi k\phi\right], \\
    \phi &= \frac{t-t_0}{P},
  \end{aligned}
  \label{eq:template}
\end{equation}
is fitted with up to $n_h=12$ harmonics, or 6 for \bcep{} stars, the order chosen by the Bayesian information criterion, together with the per-camera offsets and scales. For \bcep{} stars, which are low-amplitude and multi-periodic, and for RR~Lyrae stars observed by TESS, the residual from the dominant mode is searched for independent frequencies that are not harmonics of it; up to five (\bcep{}) or three (RR~Lyrae) are fitted as sinusoids and subtracted, down to a signal-to-noise of four in the residual spectrum (with an absolute floor of 0.2 per cent of the dominant amplitude), before any timing is attempted. This floor reaches the millimagnitude non-radial modes of first-overtone RR~Lyrae stars \citep{Netzel2024}. The test pass prewhitened only to two per cent of the dominant amplitude and left those modes in the light curve, and they are exactly what makes a per-cycle \oc{} non-white (Section~\ref{sec:topscore}). The full-frame-image rows, timed in blocks, keep the test-pass timing. RR~Lyrae stars in the ground-based surveys are not prewhitened, because nightly sampling places daily aliases where a second mode would be (Section~\ref{sec:beating}). Without this prewhitening the beat between modes is read as a large, coherent \oc{} signal (on a synthetic two-mode star the \oc{} scatter falls from $\sim$500~s to $\sim$10~s once the second mode is removed). Removed modes are recorded, and those within 15 per cent of a harmonic of the dominant mode are flagged, because a genuine phase modulation places sidebands there. Stars whose template peak-to-peak amplitude is below 0.2 per cent are not timed.

\subsection{Dense timing}
\label{sec:dense}

When a light curve carries at least 30 points per pulsation cycle (TESS two-minute data on any of our classes) each cycle is timed individually. Within the window of cycle $c$ the template is refitted with a single free time shift $\Delta_c$, holding the shape fixed and allowing an amplitude scale, by least squares against the flux with its errors; the shift and its formal error, the number of points, the reduced $\chi^2$ and the amplitude ratio are recorded. Formal errors are inflated by $\sqrt{\chi^2_\nu}$. On a bright TESS RR~Lyrae star the per-cycle precision is 10--20~s; on a TESS \bcep{} star, with its low amplitude, about 130~s against an \oc{} scatter of $\sim$800~s; on a TESS Cepheid, whose 2-min sampling puts thousands of points in each cycle, it is set by the star's own shape changes rather than by photon noise. The \oc{} series is $\Delta_c$ against $c$, and the ephemeris (a quadratic in cycle number, absorbing period error and any constant $\dot P$) is removed. Because the injected code of Section~\ref{sec:staircase} is a ramp plus steps, the ramp is absorbed here and the steps survive.

\subsection{Sparse timing and the jitter test}
\label{sec:sparse}

Ground-based surveys observe once or a few times per night, so an individual cycle of a 0.5~d RR~Lyrae star is never timed. Two statistics replace the per-cycle shift. First, each epoch's flux residual from the template is converted to an implied timing offset through the template slope at that phase, $\delta t_i = -r_i/\dot F(\phi_i)$, and these are averaged with inverse-variance weights over blocks of $K$ consecutive cycles chosen so that each block holds at least a dozen epochs ($K\approx20$ for ASAS-SN and $\approx35$ for ZTF on a typical RR~Lyrae star); the block \oc{} is then analyzed exactly as the dense series is. The conversion linearizes the template about the epoch's phase and so assumes that the offset is small compared with the scale over which $\dot F$ changes: on the steep rising branch of an RRab star, where the slope changes fastest, a large offset is read as a smaller one, which biases the block \oc{} and the jitter variance low rather than manufacturing structure. Second, and more powerfully, the squared residuals are regressed on the squared template slope,
\begin{equation}
  r_i^2 = \sigma_f^2 + \sigma_t^2\,\dot F(\phi_i)^2 ,
  \label{eq:jitter}
\end{equation}
whose slope coefficient $\sigma_t^2$ is the variance of cycle-to-cycle phase jitter, whatever its origin. A phase code that block averaging would erase still inflates $\sigma_t^2$. The jitter test requires the period to have been refined to $\sim10^{-6}$~days first, since a period error over a multi-year baseline would look like jitter. A quadratic timing drift is also fitted through the template slope and removed from the residuals before the regression. This drift is the constant $\dot P$ that the ephemeris of the dense path absorbs, and it is worth tens of seconds over a five-year baseline at a typical RR~Lyrae $\dot P$ of a few $10^{-10}$~days per day. The significance of the jitter is calibrated with $10^4$ bootstrap draws in which residuals are permuted among epochs.

\subsection{Whiteness statistics}
\label{sec:whiteness}

Every \oc{} series (dense or block) is tested for non-white structure. The statistics of this subsection read the detrended \emph{level} series; those of Section~\ref{sec:decode} read its first differences, and the two are complementary: under Equation~\ref{eq:staircase} a fair-coin code turns the level series into a random walk, whose excursions the autocorrelation statistics detect, while its steps are white and carry the bits. The autocorrelation function is computed to a lag budget of a quarter of the series, and its maximum absolute value and the Ljung--Box portmanteau statistic \citep{Ljung1978} are compared against a null of random permutations of the series. The permutation null is used deliberately in place of the $\chi^2$ tail of the Ljung--Box statistic, which is not trustworthy at the lag budget that a repeating bit word needs. The autocorrelation at prime lags, the statistic that produced the KIC~5520878 candidate of \citet{Hippke2015}, is retained as a diagnostic. The permutation entropy \citep{Bandt2002} of the series is tested the same way, jumps larger than $5\sigma$ are counted, and the periodogram of the \oc{} is searched for secondary periods, which are labeled by what they match: harmonics and $2P$/$4P$ multiples of the pulsation period, the TESS orbital and sector periods (13.7 and 27.4~d), daily, lunar and annual aliases for ground-based data, and the baseline itself. All Monte-Carlo $p$-values are add-one, $(h+1)/(N+1)$, so that a statistic that beats every draw reports $1/(N+1)$ rather than zero and the number of draws is stored beside it.

\subsection{Decode statistics}
\label{sec:decode}

The whiteness statistics test whether the \oc{} is unpredictable. A message is something different: it is \emph{decodable}, and a message carrying information does not repeat, so the autocorrelation statistics, which are powered by repetition, are the wrong instrument for a long word. Four statistics address decodability directly. All of them read the \emph{steps} of the \oc{}, the first differences between consecutive timed cycles (Equation~\ref{eq:staircase}); a step across a gap in the data is dropped rather than averaged, because the sum of unobserved bits plus whatever the gap did to the ephemeris carries no information about any single bit.

\paragraph{Fold-first description length.} For each candidate word length $N_{\rm bit}$ from 4 to $\min(64, n/2)$ and each bit period $k\in\{1,2,4,8\}$ cycles per bit, the steps are averaged per residue of $\lfloor c/k\rfloor \bmod N_{\rm bit}$, the $N_{\rm bit}$ means are split into two levels by their best two-cluster partition, and the resulting word is priced by its description length \citep{Rissanen1978}: $N_{\rm bit}\ln2$ for the bits, $\ln n$ per level, plus the entropy of the per-cycle votes that disagreed with their residue's majority. Because we fold before thresholding, a word repeated $r$ times is read at $\sqrt r$ times the per-cycle signal-to-noise ratio. This is the only way to detect a code at the kick ceiling. The minimum description length over $(N_{\rm bit},k)$ is compared against the same permutation null, with the $(N_{\rm bit},k)$ search repeated on every draw so that the look-elsewhere cost over word length is inside the $p$-value. A companion \emph{squareness} statistic, the fraction of step variance explained by the two-level word beyond the better of a sinusoid at the same period and the best sinusoid at any frequency up to the once-per-cycle Nyquist limit, is required to exceed 0.25; a permutation null is white, so a smooth Blazhko or light-travel-time modulation beats it too, and only the two-level fit separates a code from a wave.

\paragraph{Quantised steps.} Otsu's between-class share of the step variance \citep{Otsu1979} is $2/\pi$ for Gaussian steps (a normal split at its mean leaves two half-normals with means $\pm\sqrt{2/\pi}\,\sigma$, so the between-class variance is $(2/\pi)\sigma^2$) and tends to 1 for two clean levels. A coded clock produces two clean levels and a random walk does not. The statistic does not need the word to repeat, so it is the route for a long word on a short baseline. It is calibrated against the permutation null.

\paragraph{Matched filter against catalogs.} The folded step means are correlated against every window of a catalog of bit streams: the primes, Fibonacci numbers, counting numbers, powers of two, squares and triangular numbers, each packed in minimal binary and in 4- and 8-bit fixed width, and the binary expansions of $\pi$ (with and without its integer part) and $e$, computed 4096~bits deep. Every window length $12\le N_{\rm bit}\le40$ and every rotation of the word is tried, at the word's own bit period and in both polarities (Section~\ref{sec:onesided}), and each of those freedoms is charged. The statistic is the amplitude of the catalog pattern in the centered folded means divided by the noise those means carry, and the best of every trial is charged by a Bonferroni correction that is flat over the finite numeric catalogs and weighted by starting offset over the expansions. A transmitter sending $\pi$ sends it from the start, and an expansion is unbounded, so a flat count over it would measure only how many bits we chose to store; offsets $k$ instead carry a prior $w_k\propto(k+1)^{-2}$, which sums to $\zeta(2)$ and lets the expansions be searched arbitrarily deep without inflating any $p$-value, while a match that starts 40~bits into an expansion is penalized $\sim1700$ times more heavily than a match at its start. A window may disagree with the fold's own two-level word in up to $\lfloor N_{\rm bit}/10\rfloor$ bits and still name it; the tolerance is paid for in a second, coincidence $p$-value, the chance that a word with the observed number of set bits $w$ lands within $m$ flips of any catalog alignment,
\begin{equation}
  p_{\rm bits} = \min\!\left(1,\ \frac{2\,T\,R(N_{\rm bit},w',w,m)}{\binom{N_{\rm bit}}{w}}\right),
  \label{eq:pbits}
\end{equation}
with $T$ the number of alignments tried (windows at that length times rotations for the numeric catalogs; $S\,\zeta(2)(k+1)^2N_{\rm bit}$ for $S=3$ expansions at offset $k$, the inverse of the offset weight $w_k$ times the rotations, so that the union bound uses the weighted count), $w'$ the window's own number of set bits and $R$ the number of words of composition $w$ reachable within $m$ flips of a window of composition $w'$. Conditioning on the composition matters: a 28-bit word with a single set bit is exactly a window of the powers of two in some rotation, and under a fair-coin null it would be named at $p\sim10^{-3}$; conditioned, it is named at $p=1$, as a lone outlier step should be. The reported catalog $p$-value is the larger of the amplitude and coincidence $p$-values, so a claim must survive both the noise null and the coincidence null. Rejecting on any weighted $p$-value below $\alpha$ controls the family-wise error at $\alpha$ by the union bound, because the weights sum to at most one.

\paragraph{A named word.} Independently of stage 1, the same matched filter names the decoded word of every star in the store, with the disagreeing bits marked, for display and for the ranking bonus of Section~\ref{sec:score}. Of the 97{,}725 timed light curves, 1129 carry a named word at any significance, 12 at $p<0.01$ and 5 at $p<10^{-3}$. These $p$-values correct for the look-elsewhere search within one star but not for the search having been run on every star. A Benjamini--Hochberg pass \citep{Benjamini1995} at a false-discovery rate of 0.05 over all timed light curves needs the smallest $p$ below $5\times10^{-7}$. It still needs $2\times10^{-6}$ after the 72{,}265 series whose decoded word is shorter than the 12 bits any catalog window requires are dropped, since those can never be named \citep[Tarone's reduction]{Tarone1990}. The best real star is at $6.0\times10^{-5}$, and no named word survives either pass. We therefore treat the per-star gate as a screening and display quantity, not as a detection threshold. Section~\ref{sec:twostars} sets out what would count as a detection beyond that gate.

\subsection{The vetting gate}
\label{sec:triage}

A star is flagged for vetting if any of the following falls below $10^{-3}$ against its null: the Ljung--Box or maximum-autocorrelation permutation $p$-value; the jitter bootstrap $p$-value with a jitter variance at least four times its error (sparse mode); the fold-first description-length $p$-value together with the squareness condition; the quantised-steps $p$-value together with squareness; or the catalog matched-filter $p$-value together with squareness. A named word at $p<0.01$ whose catalog window beats a free sinusoid on the steps is also a route to vetting. The gates were set so that the null draws can resolve them (2000 permutations; $10^4$ for the jitter bootstrap). The three decode routes exist because the autocorrelation routes miss codes that the decoder reads: a 21-bit counting word at $\dtp=3\times10^{-3}$ reached them in 5 per cent of injections but was decoded exactly in more than half, and with the decode routes none of 80 uninjected quiet controls changed verdict (Section~\ref{sec:injection}).

\subsection{Vetting}
\label{sec:vetting}

Stage 1 only flags non-white \oc{}. For each flagged star, stage 2 tests in order whether the structure is an artifact, a shape change, red noise or systematics, or known astrophysics, and only then whether it is a code, and assigns one of the classes in Table~\ref{tab:classes}; a star that fails more than one test takes the class of the first veto in this order.

\begin{enumerate}
\item \emph{Artifact.} The \oc{} is re-measured with the cycle boundaries shifted by half a period and with a template four harmonics richer; real timing survives both (Ljung--Box $p<10^{-2}$ on each re-measurement), whereas a template-fit artifact does not. In the test pass, the sign of such an artifact was a spurious secondary period at exactly $4P$ or $2P$ in stars of very different periods.
\item \emph{Timing versus shape.} An independent timing method, the crossing time on the steepest branch of the rising light curve, must track the template \oc{} (Spearman $\rho>0.5$), and $|$\oc$|$ must not track the per-cycle $\chi^2_\nu$ or amplitude ($\rho<0.3$ or $p>10^{-3}$); a light-curve shape change leaking into the shift is the second most common failure mode after template artifacts.
\item \emph{Red noise and systematics.} A moving-block bootstrap \citep{Kunsch1989}, with block length set from the short-range correlation, tests whether the long-range autocorrelation exceeds what short-range correlation alone predicts; the series is split into independent segments (TESS sectors or observing seasons) and the structure must persist in at least two; and every secondary period is compared with the harmonics and $2P$/$4P$ multiples, the TESS orbit and sector periods, the sampling aliases of ground-based data, and the baseline.
\item \emph{Known astrophysics.} Four signatures are hard vetoes, unless the two-level code fit explains more of the variance than the sinusoid: secondary periods in the Blazhko range for RR~Lyrae stars; beats between the dominant and the prewhitened modes for \bcep{} and RR~Lyrae stars; a light-travel-time sinusoid (a Keplerian fit to the \oc{}) that dominates its variance; and a common peak in the periodograms of the per-cycle \oc{} and the per-cycle amplitude, both below a false-alarm probability of $10^{-3}$ and within $0.02$~cycles$^{-1}$ of each other. The last of these targets beating: when two modes beat, the phase and amplitude of the dominant mode are modulated together, which a trigger that only advances the next pulse would not do. A secondary period longer than a third of the baseline defines no harmonic family, so a short period is not folded into the harmonics of a secular term. Other short secondary periods are annotated rather than vetoed, because a message with a period of tens of cycles lands there too.
\item \emph{A code, and the same code twice.} The fold-first, quantised-step and catalog statistics are recomputed on the re-measured \oc{}, and each half of the baseline is decoded separately and must give the same word; repetition, not one excursion, is what the fold integrates.
\item \emph{Catalogs.} TIC magnitude and contamination ratio, the SIMBAD type and the Gaia~DR3 variability class are attached, so that a blended neighbour or a misclassified eclipsing binary is visible at a glance.
\end{enumerate}
Ground-based candidates receive the subset that applies: template refit, per-camera jitter consistency, season consistency, alias classification, and a coherence test on the per-epoch offsets $\delta t_i$. The offsets of a code with one bit per cycle are white; those of a Blazhko star are coherent over the modulation period. One slow sinusoid (10--1000~d, capped at half the baseline) is fitted to the $\delta t_i$ together with the quadratic ephemeris, If the sinusoid explains more than half of the star's excess jitter variance, with a $\chi^2$ improvement above 30 and a period away from the sampling aliases, the star is classed ASTRO. A repeated word is itself coherent at the word period, so before the veto fires the offsets are folded on the fitted period, one bin per bit at each bit period the decoder searches, and the steps are priced as a two-level code (ln~2 per bit) against the best wave; a fold whose two-level code wins by five nats or more is not vetoed. On synthetic sparse series the test vetoes Blazhko sinusoids of 15--120~d and leaves words held for 8--32 cycles per bit, which win by 50--200 nats.

\subsection{Beating at an irrational period ratio}
\label{sec:beating}

The KIC~5520878 episode (Section~\ref{sec:hippke}) is the specific failure that this search has to avoid, so we tested for it directly rather than assuming the vetting would catch it. A second pulsation mode sampled once per cycle of the first appears in the \oc{} as $x_c = A\sin(2\pi\nu c + \phi)$ with $\nu = P_1/P_2 \bmod 1$ irrational; for KIC~5520878, $\nu\approx0.618$, whose continued-fraction convergents are the Fibonacci ratios. A series of this kind nearly repeats when it is folded on any convergent denominator $q$, because $\nu q$ is within $1/q$ of an integer, so every residue collects samples of almost the same phase and the fold is a clean sinusoid over $q$ residues. A two-level split of that fold is a stable $q$-bit ``word.'' For a pure two-mode series with $\nu=0.618$ at three to ten times the noise, the fold on a convergent denominator gives a 21-bit word, and for $\nu=\sqrt2$ a 29-bit word. The whiteness statistics flag the series. Against a sinusoid at the word's own period the two-level fit has squareness $0.76$--$0.79$, because the beat is a sinusoid at the aliased period $1/(1-\nu)\approx2.6$ cycles that merely folds well at that length. The squareness statistic uses the best sinusoid at any frequency and reads zero, while the injected staircase code of the synthetic RRab control of Figure~\ref{fig:synthetic} keeps its squareness at $0.93$ (the primes word of the synthetic Cepheid control, $0.87$). The matched filter does not name either word. A word from an irrational rotation has at most $2m$ distinct substrings of each length $m$, against the $2^m$ a message can use, and we reject any word of 21 or more bits that meets the bound; a fair-coin word of that length meets it with probability of about one per cent.

The test that matters is the star itself. KIC~5520878 is in the target lists of Section~\ref{sec:surveys} under its TESS, ASAS-SN and ZTF names, but TESS holds only two 30-minute full-frame sectors for it and the ground-based rows average tens of cycles per block, so none of them can resolve a beat that turns over every 2.6 cycles. The pipeline therefore accepts Kepler as a second dense survey for named targets, with the Kepler month, quarter and year in place of the TESS orbit and sector as sampling periods, and the 19 short-cadence months of KIC~5520878 (723{,}000 points, 395 per cycle, 1841 timed cycles over 856~d) were run through the same stage-1 and stage-2 cuts as every other star. Prewhitening removed one independent mode, at $P_1=0.17010$~d ($f_1/f_0=1.582$) and 2.4 per cent of the peak-to-peak amplitude, just above the floor. The \oc{} that remained is not white: the Ljung--Box test and the permutation entropy both flag it, and its strongest secondary period, 4.79 cycles, is the subharmonic of the aliased residual beat. The decode routes do not read it as a message: the two-level word is 24 bits long, its squareness against the best free sinusoid is $0.00$, it matches no catalog, and the sequence $p$-value is unity, so the star reaches triage as a multi-periodic pulsator, not as a code candidate. Stage-2 vetting of the same series classifies the star as ASTRO: the 4.79-cycle secondary period is, to within a per cent, the period of the free sinusoid that explains the decoded word's steps as well as two levels do (its frequency is $0.209$ per cycle), which is the hard veto of Section~\ref{sec:vetting} for a second mode sampled once per cycle. The star is displayed in the web interface (Section~\ref{sec:viewer}) alongside its TESS, ASAS-SN and ZTF rows, none of which reaches vetting as anything but a quiet or secular clock.

\begin{table}[tp]
\centering
\caption{Vetting classes and their base scores \label{tab:classes}}
\small
\fittable{\begin{tabular}{llr}
\toprule
Class & Meaning & Base \\
\midrule
SURVIVOR   & survived every stage-2 veto                  & 0.70 \\
WEAK       & non-white \oc{}, not independently confirmed & 0.45 \\
TRIAGE     & stage-1 flag, not yet vetted                 & 0.32 \\
ASTRO      & explained by known astrophysics              & 0.12 \\
RED NOISE  & short-range correlated noise                 & 0.10 \\
P-MULTIPLE & $2P$/$4P$ sampling or period doubling        & 0.08 \\
SYSTEMATICS& survey orbit, sector or sampling period      & 0.07 \\
SHAPE      & light-curve shape change, not timing         & 0.06 \\
QUIET      & no stage-1 flag                              & 0.06 \\
ARTIFACT   & pipeline artifact                            & 0.04 \\
\bottomrule
\end{tabular}}
\end{table}

\subsection{Ranking}
\label{sec:score}

Every star receives a score, which the web interface uses to sort the list. The stage-2 class sets a base score (Table~\ref{tab:classes}), and several terms are added to it. The evidence term is the mean, over the available statistics (Ljung--Box, maximum autocorrelation, block bootstrap, description length, quantised steps, catalog match), of a log-evidence curve that is zero at $p\ge0.05$ and saturates at $p=10^{-6}$, multiplied by a signal-to-noise factor $\min(1,\ \sigma_{\rm O-C}/3\bar\sigma_t)$. A robustness term credits survival of the boundary shift, the richer template, the independent timing and segment consistency. Each unexplained secondary period is penalized, and so is sparse (block-averaged) timing, which is intrinsically weaker evidence than per-cycle timing. A named catalog word at $p<0.01$ earns a bonus of up to 0.20, on the same evidence curve. Vetoed classes are capped below every unvetoed one, and a sparse survivor cannot outrank a dense one. The score is a ranking device for human attention, not a probability.

\subsection{Sensitivity by injection}
\label{sec:injection}

\begin{figure*}[tp]
  \centering
  \includegraphics[width=\textwidth]{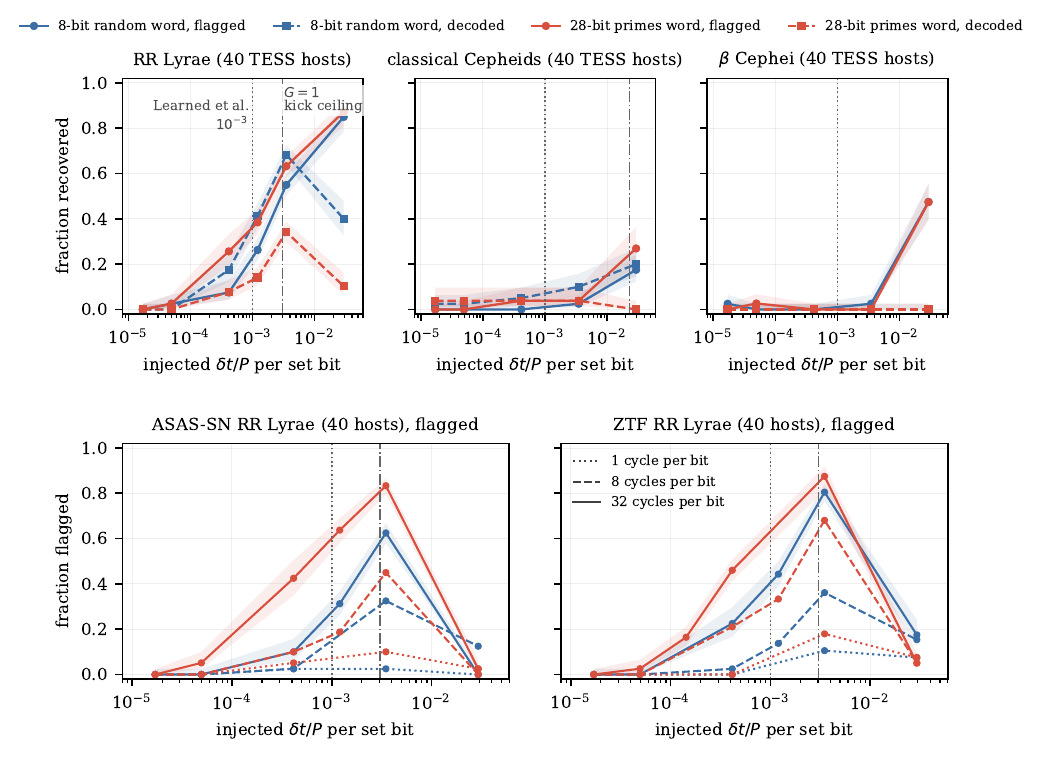}
  \caption{Recovery of injected phase codes on real, uninjected-quiet hosts, cumulative (staircase) waveform, as a function of the advance per set bit. Top: TESS, one bit per cycle, per class. Solid lines: the star reaches the vetting gate (flagged). Dashed: at least 90 per cent of the injected bits are read back correctly. Blue: an 8-bit random word; red: the 28-bit primes word. Bottom: the flagged fraction on ASAS-SN and ZTF RR~Lyrae hosts, with the code held for one (dotted), eight (dashed) and 32 (solid) cycles per bit; no ground-based injection was decoded. Shaded bands are one-sigma Wilson intervals. Dotted vertical line: the $\dtp=10^{-3}$ scaling of \citet{Learned2008}; dash-dotted: the $G=1$ kick ceiling of a storage-free transmitter with a complete sphere at unit efficiency, per class (Section~\ref{sec:swarm}): $3\times10^{-3}$ for the RR~Lyrae stars and $0.023$ for the Cepheids, at the amplitude where the injection stops being a phase code; the \bcep{} ceiling, $0.064$, lies beyond the axis. In the RR~Lyrae panel the 8-bit word is decoded more often than it is flagged at $3.5\times10^{-3}$, because the decode routes promote on $p<10^{-3}$ with squareness rather than on bit recall. The RR~Lyrae arm is the only one that reaches half completeness below $\dtp=10^{-2}$. The fall of every curve at the largest amplitude marks where the injection starts to drag the template and period with it. Produced by \texttt{figures/make\_completeness.py} from the released injection database.}
  \label{fig:completeness}
\end{figure*}

A null result is a measurement only if the search's sensitivity is known, so the pipeline includes an injection stage that writes a phase code of known amplitude into \emph{real} light curves and pushes them through the production stage-1 path unchanged. The injection is a relabelling of the time stamps, $t'=t-\delta t\sum_{k\le c(t)}b_k$ (Equation~\ref{eq:staircase}), so the flux, noise, cadence, gaps and whatever astrophysics the host carries are the star's own; a fully synthetic control would measure the detector against the idealization it assumes, whereas this measures it against the sky. The hosts are stars that stage 1 classed as quiet, and every host is also run with nothing injected. These uninjected runs confirm that the quiet verdicts are reproducible under the injection machinery: none of the 80--160 uninjected controls per class in the main runs reached the vetting gate. Because the hosts were pre-selected as quiet, this is a reproducibility check and not a false-alarm rate, which is treated below. The uninjected runs also give the chance decode rate: none of 1631 uninjected decodes reached 90 per cent recall of the word that would have been injected. The amplitude is searched rather than sampled: a log-spaced scan brackets the 50 per cent crossing, a refine pass is placed inside the bracket, and a logistic completeness curve is fitted to every injection by maximum likelihood. A class that never reaches the target rate below $\dtp=0.05$, where the shift is a twentieth of a cycle and the template follows it, is recorded as a lower bound.

Figure~\ref{fig:completeness} shows the result for 40 TESS hosts per class and Table~\ref{tab:sensitivity} the fitted thresholds. Three findings shape the rest of the paper. First, the RR~Lyrae stars are the only population in which this search can detect a code at a credible amplitude: an 8-bit word is flagged half the time at $\dtp\approx4\times10^{-3}$ and the 28-bit primes word at $\approx2.4\times10^{-3}$, within a factor of a few of the \citet{Learned2008} scaling and at the $G=1$ kick ceiling of a fully enclosed star. The Cepheid arm, whose median TESS host carries only 18 timed cycles, and the \bcep{} arm do not reach half completeness at any amplitude that is still a phase code; the scatter ratios behind that are those of Section~\ref{sec:foreground}. Second, recovery is set by how many cycles were timed: at fixed amplitude, RR~Lyrae hosts with more than 60 timed cycles are found every time while those with fewer than 25 are missed nine times in ten, which is a coverage problem that more sectors and cross-survey stitching address, not a statistics problem. Third, the decoder often reads a word at amplitudes where the autocorrelation gates would never have looked at the star, which is why the decode routes of Section~\ref{sec:triage} exist.

The ground-based arm, which holds 82 per cent of the light curves, was measured separately on 40 quiet ASAS-SN and 40 quiet ZTF RR~Lyrae hosts, at one, eight and 32 cycles per bit (Table~\ref{tab:sensitivity}, lower rows; Figure~\ref{fig:completeness}, lower panels). A one-bit-per-cycle code is invisible there at any amplitude that is still a phase code: block averaging sums tens of bits into every step, and the flag rate never exceeds 18 per cent below $\dtp=0.05$. A code held for 32 cycles per bit, about 18~d for a 0.57~d star, is flagged half the time at $\dtp\approx(0.6$--$2.6)\times10^{-3}$, through the jitter statistic and the per-cycle epoch fold of Section~\ref{sec:sparse}, and one held for eight cycles at $(2$--$6)\times10^{-3}$; but no sparse injection at any bit period was decoded or named. Stage~2 was run on every flagged ZTF injection. Of the 397 flagged below the ceiling amplitude it classed 86 SURVIVOR, 36 WEAK, 2 SYSTEMATICS and 1 RED NOISE, rejected 159 as artifacts (the block \oc{} or the jitter did not survive a richer template or shifted block boundaries) and 113 on the astrophysics vetoes, 43 of those with the coherence test of Section~\ref{sec:vetting} among the reasons. Near the flag threshold the SURVIVOR fraction is a fifth to a third for the 8-bit word at either bit period and for the primes word at eight cycles per bit, and a few per cent for the primes word at 32 cycles per bit, whose 500-day staircase appears in the block \oc{} as a secondary period in the Blazhko range or as an orbital sinusoid. The sparse arm's completeness at the SURVIVOR gate is therefore below one half at every amplitude, and the ground-based rows of Table~\ref{tab:sensitivity} are stage-1 numbers. The ground-based surveys can detect a slow code as timing structure but cannot decode it, and the Gaia~DR4 extension of Section~\ref{sec:catalogs} would share both limitations.

Four further controls cover cases that Figure~\ref{fig:completeness} does not.

\paragraph{Hosts with their own structure.}
The completeness above is conditional on the host being quiet; a code in a Blazhko star or a beating \bcep{} star is not represented. We drew forty TESS RR~Lyrae hosts without regard to their stage-1 verdict. With nothing injected, 60 per cent of them were flagged under the test-pass prewhitening, and 25 per cent under the noise floor of Section~\ref{sec:template}. In these hosts an injected 8-bit word is decoded 65 per cent of the time at $\dtp=2.4\times10^{-3}$, and the primes word 31--41 per cent of the time. At the decode gate, then, a code in a star with structure of its own is not much harder to read than one in a quiet star. The flag gate gives no information here, because most of those hosts were flagged already.

\paragraph{The survey flag rate.}
The false-positive rate of stage~1 is the flag rate of the survey itself, an upper bound that is exact under the null that no star carries a beacon: 39 per cent overall (Table~\ref{tab:targets}): 7, 21 and 57 per cent of the TESS Cepheids, RR~Lyrae and \bcep{} stars, 14, 46 and 12 per cent of the ASAS-SN ones, and 24 and 44 per cent of the ZTF Cepheids and RR~Lyrae stars, against gates set at $p<10^{-3}$. The 6823 survivors of Section~\ref{sec:candidates} should be read against that number.

\paragraph{A two-mode beat.}
The two-mode beat control of Section~\ref{sec:beating} was run on the same 40 quiet TESS RR~Lyrae hosts, with a second mode at $\nu=0.618$ and $0.707$ written into the clock at three to ten times the \oc{} scatter. It reaches the flag gate in 74--94 per cent of injections, as a non-white clock should, but none of 228 injections passes the squareness guard or is promoted by any decode route, none is decoded as a word, and 8 per cent are named by the catalog matcher at some $p$, none below the 0.01 gate.

\paragraph{The thermal tail.}
With each set bit's advance accruing as $1-e^{-(n+1)/\tau}$ over the following cycles at $\tau=1$ (Section~\ref{sec:staircase}), the 8-bit word on 40 quiet TESS RR~Lyrae hosts is flagged half the time at $\dtp=1.1\times10^{-2}$ rather than $4.0\times10^{-3}$, decoded at $5.8\times10^{-3}$ rather than $4.2\times10^{-3}$, and promoted by the quantised-step route at $3.6\times10^{-2}$ rather than $2.6\times10^{-2}$ (Table~\ref{tab:sensitivity}). A one-cycle tail costs the flag gate a factor of 2.6 and the decode gate 1.4, a penalty of the same order as the staircase itself, so the two-level discriminant survives the physics. Six of that run's 80 uninjected controls reached the flag gate, against none in the earlier runs; its hosts were drawn as quiet from the store after the re-timing of Section~\ref{sec:template}, and a quiet verdict is reproducible only under the pipeline that gave it.

\begin{table*}[tp]
\centering
\caption{Amplitude for 50 per cent recovery, staircase waveform \label{tab:sensitivity}}
\fittable{\begin{tabular}{lllcc}
\toprule
Survey, class & Word & $k$ & Flagged (\dtp) & Decoded (\dtp) \\
\midrule
TESS RR Lyrae    & 8-bit random  & 1  & $4.0\times10^{-3}$ & $4.2\times10^{-3}$ \\
TESS RR Lyrae    & 28-bit primes & 1  & $2.4\times10^{-3}$ & $\ge0.05$ \\
TESS RR Lyrae, one-cycle tail & 8-bit random & 1 & $1.1\times10^{-2}$ & $5.8\times10^{-3}$ \\
TESS Cepheid     & 8-bit random  & 1  & $\ge0.05$ & $\ge0.05$ \\
TESS Cepheid     & 28-bit primes & 1  & $\ge0.05$ & $\ge0.05$ \\
TESS \bcep{}     & 8-bit random  & 1  & $\ge0.05$ & $\ge0.05$ \\
TESS \bcep{}     & 28-bit primes & 1  & $\ge0.05$ & $\ge0.05$ \\
ASAS-SN RR Lyrae & 8-bit random  & 1  & $\ge0.05$ & $\ge0.05$ \\
ASAS-SN RR Lyrae & 28-bit primes & 1  & $\ge0.05$ & $\ge0.05$ \\
ASAS-SN RR Lyrae & 8-bit random  & 8  & $\ge6\times10^{-3}$ & $\ge0.05$ \\
ASAS-SN RR Lyrae & 28-bit primes & 8  & $4.4\times10^{-3}$ & $\ge0.05$ \\
ASAS-SN RR Lyrae & 8-bit random  & 32 & $2.6\times10^{-3}$ & $\ge0.05$ \\
ASAS-SN RR Lyrae & 28-bit primes & 32 & $8.9\times10^{-4}$ & $\ge0.05$ \\
ZTF RR Lyrae     & 8-bit random  & 1  & $\ge0.05$ & $\ge0.05$ \\
ZTF RR Lyrae     & 28-bit primes & 1  & $\ge0.05$ & $\ge0.05$ \\
ZTF RR Lyrae     & 8-bit random  & 8  & $5.7\times10^{-3}$ & $\ge0.05$ \\
ZTF RR Lyrae     & 28-bit primes & 8  & $2.4\times10^{-3}$ & $\ge0.05$ \\
ZTF RR Lyrae     & 8-bit random  & 32 & $1.5\times10^{-3}$ & $\ge0.05$ \\
ZTF RR Lyrae     & 28-bit primes & 32 & $6.0\times10^{-4}$ & $\ge0.05$ \\
\bottomrule
\end{tabular}}

\begin{minipage}{\textwidth}
\vspace{1pt}
\small \textsc{Note}---$k$ is the number of pulsation cycles per bit. 40 hosts per row; 130--360 injections per TESS row, 160--320 per ground-based row. A ``$\ge$'' entry means the target rate was not reached below the amplitude at which the injection stops being a phase code (0.05), or, for the one entry at $6\times10^{-3}$, below the largest amplitude the scan placed short of it. The decoded gate for the primes word requires 28 consecutive correctly read bits and is therefore harsher than the flagged gate, which the fold integrates over repeats. TESS rows measured 2026 September 17 with the fold decoder and the one-cycle-tail row (Section~\ref{sec:staircase}) on 2026 September 25; ASAS-SN rows 2026 September 24 with the kick and epoch-fold routes; ZTF rows re-measured on 2026 September 25 with the quadratic drift removed before the jitter regression (Section~\ref{sec:sparse}) and stage~2 run on every flagged injection, which moved no ZTF crossing by more than 10 per cent. The ground-based entries are the raw 50 per cent crossings between adjacent amplitudes, because a logistic fit that includes the ceiling amplitude, at which the template follows the code and the flag rate collapses, is bent by it. The two-mode beat control of Section~\ref{sec:beating} is described in the text.
\end{minipage}
\end{table*}

\begin{figure*}[tp]
  \centering
  \includegraphics[width=\textwidth]{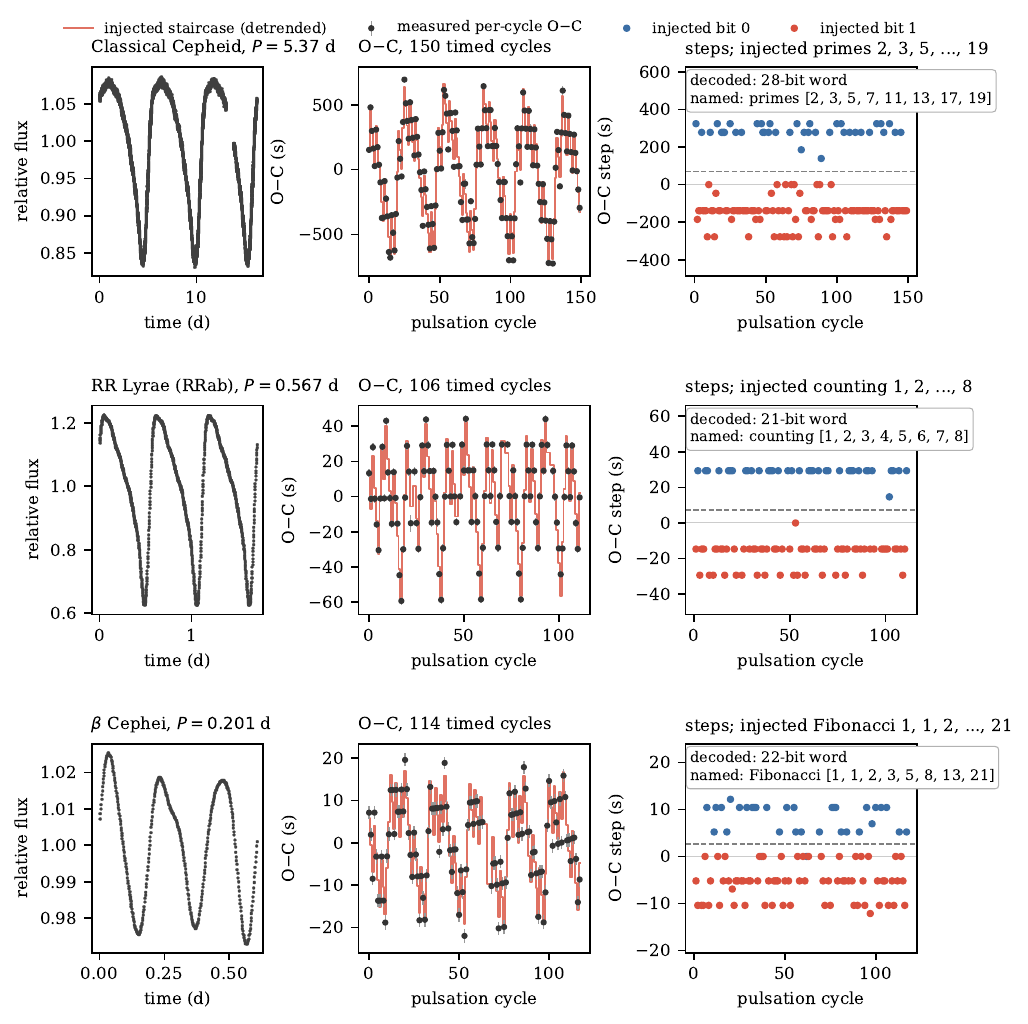}
  \caption{The three synthetic controls timed by the production stage-1 code. Rows: a classical Cepheid carrying the primes 2--19 in binary, an RRab star carrying the counting numbers 1--8, and a two-mode \bcep{} star carrying the Fibonacci numbers 1--21, each at $\dtp=10^{-3}$ per set bit, one bit per cycle, TESS two-minute sampling with sector gaps. Left: three cycles of the light curve. Middle: the per-cycle \oc{} from the template time-shift fit with the injected staircase drawn through it; both are detrended, since the period search absorbs the mean advance. Right: the first difference of the \oc{}, the series the decoder reads, colored by the injected bit; a clear bit reads as $+d\,\delta t$ and a set bit as $-(1-d)\,\delta t$ once the ramp is removed. Where the timer's cycle windows do not coincide with the transmitter's, a window straddling two bits reads an intermediate step (visible in the Cepheid row); the fold-first decoder recovers the word regardless. Produced by \texttt{figures/make\_synthetic.py}.}
  \label{fig:synthetic}
\end{figure*}

Figure~\ref{fig:synthetic} shows what a detection would look like. The three built-in synthetic controls of the pipeline (a Cepheid, an RRab and a two-mode \bcep{} star at TESS two-minute cadence with sector gaps) each carry a different recognizable word at $\dtp=10^{-3}$: the primes, the counting numbers and the Fibonacci numbers. Timed by the production code, their \oc{} series show the cumulative staircase of Equation~\ref{eq:staircase} with its ramp absorbed into the period, their steps fall into two levels, and the decoder names all three words.

\section{The Web Interface}
\label{sec:viewer}

Every result of the pipeline is published through a browsable interface at \url{https://starseti.astrocrash.net}, the same viewer that the author uses to inspect candidates. It serves the SQLite store read-only and lists every star in the order of the score of Section~\ref{sec:score}, filterable by class, survey, vetting class and decoded sequence. For each star it draws the folded light curve and template, the \oc{} series with its stage-2 re-measurements, the autocorrelation against its null, and the decoded word with the catalog window that it was matched to and both $p$-values of Section~\ref{sec:decode}. The three synthetic controls of Figure~\ref{fig:synthetic} are pinned first, so that a real star can always be compared with what a detection would look like. The interface is meant to make the search checkable: any reader can open the highest-ranked star, see why it is ranked there and which tests it passed and failed, and see the raw timing series from which every statistic was computed. At the time of writing, the store behind it holds 97{,}725 timed light curves of 77{,}596 stars and 39{,}124 vetting results. The vetting results outnumber the flagged rows because they include controls, re-vets and rows that a later re-timing removed from triage. The store will be refreshed as the search proceeds.

\section{Candidates}
\label{sec:candidates}

Of the 38{,}230 flagged light curves, stage~2 classes 17{,}313 as ARTIFACT, 5525 as ASTRO, 5535 as WEAK, 1981 as RED NOISE, 509 as SHAPE, 307 as SYSTEMATICS and 218 as P-MULTIPLE; 19 failed to vet. The remaining 6823 are SURVIVORs, stars whose timing structure no veto explains. Nearly all of them are ground-based: 5979 from ZTF, 639 from ASAS-SN and 197 from TESS full-frame images too sparse to time per cycle. Only eight have per-cycle timing. Most of the ground-based survivors reach that class on excess per-epoch jitter alone. That is the signature a one-bit-per-cycle code would leave, but it is also the signature Blazhko modulation leaves, and the coherence test of Section~\ref{sec:vetting} is the only test on the sparse path that can tell them apart. Re-vetting the 15{,}798 ground-based rows that the test pass had classed SURVIVOR or WEAK, with the quadratic drift removed before the jitter regression and the coherence test in place, moved 2786 to ARTIFACT and 820 to ASTRO, 785 of them on the coherence test. A SURVIVOR is a star whose timing the search could not explain, not a star that it found to be artificial.

Table~\ref{tab:subtype} splits the RR~Lyrae rows by Gaia subtype, which is the cheapest test of what the sparse survivors are. The flag rate is nearly the same for RRab and RRc stars in every survey, so what stage~1 flags is not, at that stage, the Blazhko effect. The survivor rate is five to nine times higher for RRab stars than for RRc stars, in step with the Blazhko incidence of 40--50 per cent among RRab stars \citep{Prudil2017, Kovacs2020, Skarka2020} against a few per cent among RRc stars \citep{Netzel2018}, and with the lower limit of 11 per cent that a strict visual selection finds in ZTF field stars \citep{Donev2025}; the RRc stars instead go to ASTRO two to three times more often, through the periods of additional modes, which the astrophysics vetoes catch. What survives the sparse vetting is Blazhko-shaped even though what is flagged is not. We inspected the top of the list by hand in two orderings: by the score of Section~\ref{sec:score}, and by the significance of a named word.

\begin{table}[tp]
\centering
\caption{RR~Lyrae rows by Gaia subtype \label{tab:subtype}}
\small
\fittable{\begin{tabular}{llrrrr}
\toprule
Survey & Type & Timed & Flagged & SURVIVOR & ASTRO \\
\midrule
TESS    & RRab & 8807  & 22\% & 2.1\% & 2.5\% \\
TESS    & RRc  & 4007  & 18\% & 0.3\% & 4.9\% \\
ASAS-SN & RRab & 10036 & 50\% & 6.3\% & 3.9\% \\
ASAS-SN & RRc  & 4970  & 40\% & 0.7\% & 11.9\% \\
ZTF     & RRab & 42192 & 45\% & 12.8\% & 3.7\% \\
ZTF     & RRc  & 18702 & 40\% & 2.3\% & 10.9\% \\
\bottomrule
\end{tabular}}

\begin{minipage}{\columnwidth}
\vspace{1pt}
\small \textsc{Note}---Percentages are of the timed rows. The 720 RRd rows are omitted.
\end{minipage}
\end{table}

\subsection{The highest-scoring survivors}
\label{sec:topscore}

\begin{table*}[tp]
\centering
\caption{The ten highest-scoring stars \label{tab:topscore}}
\footnotesize
\fittable{\begin{tabular}{llrrrccl}
\toprule
Star & Type & $P$ (d) & $N_{\rm cyc}$ & $\sigma_{\rm O-C}$ (s) & $\nu_{\rm O-C}$ & $\nu_{\rm amp}$ & Explanation \\
\midrule
\textbf{TIC 152825245} & RRc & 0.3131 & 130 & 83 & 0.185 & 0.189 & 0.61 alias subharmonic \\
TIC 240268877 (BD+46 3337) & BCEP? & 0.3563 & 293 & 2586 & 0.162 & 0.008 & comparable modes \\
TIC 10931356 & RRab & 0.5481 & 84 & 135 & 0.311 & 0.008 & unexplained \\
TIC 467480241 & BCEP & 0.1421 & 852 & 212 & secular & 0.042 & beat, 3.4~d \\
TIC 114228663 & BCEP & 0.1581 & 247 & 665 & 0.193 & 0.101 & beat, 3.2~d \\
TIC 142240683 & RRc & 0.3113 & 135 & 21 & 0.379 & 0.002 & alias of the prewhitened 0.61 mode \\
ASASSN 360777681701 & RRc & 0.3373 & 68 & 4336 & \ldots & \ldots & sparse jitter, 4100~s \\
ASASSN 360777954158 & RRc & 0.3451 & 60 & 1605 & \ldots & \ldots & sparse jitter, 2400~s \\
ASASSN 412317075339 & RRc & 0.3417 & 59 & 2061 & \ldots & \ldots & sparse jitter, 2300~s \\
ASASSN 618475347825 & RRc & 0.3564 & 121 & 1057 & \ldots & \ldots & sparse jitter, 1050~s \\
\bottomrule
\end{tabular}}

\begin{minipage}{\textwidth}
\vspace{1pt}
\small \textsc{Note}---Classed SURVIVOR, in ranked order: the six highest-scoring of the eight TESS stars with per-cycle timing, then the first four sparse survivors, since a sparse survivor cannot outrank a dense one. Type is the Gaia~DR3 class; the query mark on BD+46~3337 is Gaia's, and the text argues the star is SPB-like. $N_{\rm cyc}$ is the number of timed cycles, or of timed blocks of 40--80 cycles for the ASAS-SN rows, and $\sigma_{\rm O-C}$ the \oc{} scatter. $\nu_{\rm O-C}$ and $\nu_{\rm amp}$ are the strongest non-secular peaks, in cycles$^{-1}$, of the Lomb--Scargle periodograms of the per-cycle \oc{} and of the per-cycle amplitude; for TIC~467480241 the strongest \oc{} peak is a secular term, and the sparse rows have no per-cycle amplitude. None of the ten decodes to a named word ($p=1$), and none has a decoded word with squareness above 0.08. The star in bold is shown in Figure~\ref{fig:candidates}.
\end{minipage}
\end{table*}

Table~\ref{tab:topscore} lists the ten stars at the top of the ranking: six TESS stars with per-cycle timing and, since a sparse survivor cannot outrank a dense one, the first four sparse ones. All share the same minimum $p$-value, the $1/(N+1)$ floor of the 2000-draw permutation null, so their order among themselves is mostly set by tie-breaking terms. None carries a named word, and none of the designed-signal discriminants of Section~\ref{sec:decode} fires. Squareness is at most 0.08 against a gate of 0.25, and no decoded word is reproduced when the two halves of the baseline are decoded separately.

Five of the six per-cycle stars have a natural explanation visible in the data: each is a residue of the modes that prewhitening removed. For the two RRc stars the \oc{} periodogram peaks at the per-cycle alias of a mode that is no longer in the light curve: TIC~142240683 at $\nu=0.379$, the alias of a 0.61 mode at $P_x=0.192$~d ($P_x/P_1=0.616$) removed at 0.3 per cent of the light curve, and TIC~152825245 at 0.185, the subharmonic of the same alias for its 0.1925~d mode, with the amplitude moving at the same frequency. It is the alias, not the mode, that appears in the timing. A mode removed from the flux to a part in a few hundred still leaves its alias in a per-cycle clock that is measured to a part in $10^{4}$. The three \bcep{} stars are beats: TIC~114228663 has \oc{} and amplitude modulated at harmonics of a 3.2~d beat between the dominant mode and one that prewhitening left unresolved, TIC~467480241 an amplitude modulation at its 3.4~d beat under a secular \oc{} term, and BD+46~3337 is the SPB-like star discussed below. The sixth, TIC~10931356, an RRab star with a 3.2-cycle \oc{} term and a flat amplitude, has no explanation in the data we hold. The four sparse survivors are ASAS-SN RRc stars whose per-epoch jitter is 1000--4000~s, 4--14 per cent of the period, on 59--121 blocks of 40--80 cycles, with a secular term as their only secondary period. A shift that large on a first-overtone star is more likely a period alias or a blended neighbour than a code, and the sparse path has no test that can distinguish these cases.

The test pass of this search, which prewhitened RR~Lyrae light curves only to two per cent of the dominant amplitude, ranked ten different TESS stars here: five RRc stars carrying the additional modes that space photometry finds in most first-overtone pulsators \citep{Moskalik2015, Netzel2019, Netzel2023, Netzel2024}, four beating \bcep{} stars, and BD+46~3337. In eight of the ten the \oc{} and amplitude periodograms peaked at the same frequency, which the zero-lag shape test of Section~\ref{sec:vetting} cannot detect, because under beating the two modulations are roughly in quadrature. All ten are now vetoed, by the noise-based prewhitening floor of Section~\ref{sec:template} and by the common-frequency veto of Section~\ref{sec:vetting}, which accounts for 493 ASTRO rows in the store. What remains at the top is what those changes cannot reach: aliases of prewhitened modes, beats the baseline cannot resolve, and sparse jitter. Two of the eight dense survivors, TIC~235416406, a \bcep{} star, and TIC~347799445, an RRc star, are unresolved beats whose \oc{} and amplitude do not peak together, so the common-frequency veto lets them through.

BD+46~3337 is not a \bcep{} star in any useful sense. Its TIC effective temperature is 10{,}400~K, far below the class, and its 8.6~h period is longer than the 2--7~h typical of it. The timed mode has a semi-amplitude of only 0.17 per cent. Three of the five modes removed before timing are within 20 per cent of that amplitude, and one is within 10 per cent of its frequency. Its \oc{} periodogram is flat. It is a low-amplitude multiperiodic star, probably SPB-like, whose ``pulsation period'' is one of several comparable modes, and its timing measures their interference.

\subsection{The most significant named words}
\label{sec:topseq}

\begin{table*}[tp]
\centering
\caption{The ten real stars with the most significant named words \label{tab:topseq}}
\fittable{\begin{tabular}{lllclccl}
\toprule
Star & Type & Named window & Encoding & $N_{\rm bit}$ & $m$ & $p$ & Stage 2 \\
\midrule
TIC 151932981 (SX Pup) & RRab & counting 73--76, inverted  & 8-bit   & 38 & 1 & $6.0\times10^{-5}$ & ARTIFACT \\
TIC 327061092         & RRc  & counting 72--75            & 8-bit   & 33 & 3 & $2.1\times10^{-4}$ & QUIET \\
ZTF 685111300023197   & RRab & triangular 561, 595, inverted & minimal & 29 & 2 & $4.9\times10^{-4}$ & ASTRO \\
TIC 379933456         & RRab & counting 73--75, inverted  & 8-bit   & 29 & 1 & $5.8\times10^{-4}$ & ASTRO \\
ASASSN 575526037701   & RRc  & counting 43--46            & minimal & 29 & 2 & $6.8\times10^{-4}$ & ARTIFACT \\
\textbf{TIC 147576267} & RRc  & triangular 561, 595, inverted & minimal & 28 & 2 & $1.3\times10^{-3}$ & QUIET \\
ZTF 333102100073787   & RRc  & $e$, inverted              & stream  & 21 & 1 & $1.4\times10^{-3}$ & WEAK \\
TIC 141590160         & RRc  & counting 72--75, inverted  & 8-bit   & 37 & 2 & $1.99\times10^{-3}$ & ASTRO \\
TIC 204639317         & BCEP & counting 50--51            & 8-bit   & 27 & 2 & $2.05\times10^{-3}$ & ARTIFACT \\
ZTF 865113100007054   & RRc  & squares 49, 64, inverted   & 8-bit   & 28 & 1 & $7.0\times10^{-3}$ & WEAK \\
\bottomrule
\end{tabular}}

\begin{minipage}{\textwidth}
\vspace{1pt}
\small \textsc{Note}---$N_{\rm bit}$ is the word length in bits, $m$ the number of bits in which the fold's own word disagrees with the catalog window, and $p$ the larger of the amplitude and coincidence $p$-values of Section~\ref{sec:decode}, corrected for the search within one star but not across stars. ``Inverted'' means the window matched with the opposite polarity. QUIET rows were not promoted by stage~1, because their windows did not beat a free sinusoid on the steps. The synthetic primes control, at $p=2.01\times10^{-3}$, would rank ninth, between TIC~141590160 and TIC~204639317. The star in bold, the first TESS star outside the counting windows, is shown in Figure~\ref{fig:candidates}.
\end{minipage}
\end{table*}

Table~\ref{tab:topseq} lists the ten real stars whose decoded words are named at the lowest $p$. None survives the across-star correction of Section~\ref{sec:decode}. Six of the ten are vetoed by stage~2, two are WEAK, and two were never promoted. Every word with a measured squareness has a value between 0.00 and 0.15, where the synthetic primes control has 0.87. On closer inspection, the named words are one astrophysical signal that happens to match one small region of the catalog.

Six of the ten are ``counting'' windows, and four of those are overlapping windows of the integers 72--76 in 8-bit fixed width, \texttt{01001000}\,\allowbreak\texttt{01001001}\,\allowbreak\texttt{01001010}\,\allowbreak\ldots. That stretch of the catalog is almost periodic with period three (\texttt{010\,010\,01\ldots}), and the decoded words are period-three repetitions such as \texttt{110110110101101}. In the test pass the three RRc stars carrying that window had \oc{} periodograms peaking at 0.36--0.38 cycles$^{-1}$, a period of 2.7 cycles, the per-cycle alias of the 0.61 mode; once those modes are prewhitened, the star that headed the list, TIC~76305643, folds into a 41-bit word that matches no catalog. The same window is still named in 91 light curves of unrelated stars across all three surveys (8 per cent of the 1129 named words), 59 of them RRc stars; a counting window starting at 40--44 is named in 95. A beacon would not send the same five integers from 91 stars, but a bias in the matcher would produce that pattern. The coincidence $p$-value (Equation~\ref{eq:pbits}) conditions on the number of set bits but treats every arrangement of them as equally likely. The aliased sinusoid of a second mode produces near-periodic arrangements far more often than that null allows, and a few catalog windows are rich in exactly those arrangements. So $p_{\rm bits}$ is too small for the most common natural signal in the sample. A null drawn from sinusoidal words at the observed alias frequency would remove this family.

A catalog-level control points the same way. In the test pass we re-ran the matcher with every catalog bit-flipped, and again with every catalog reversed, over the 12{,}054 stars with score $\ge0.5$ and a decodable series. The real catalog named 98 of them, the flipped catalog 91 and the reversed catalog 99. Sequences no transmitter would choose are named as often as the ones a transmitter would, which is the behavior of a matcher finding chance resemblances, not messages.

\subsection{Two stars in detail}
\label{sec:twostars}

\begin{figure*}[tp]
  \centering
  \includegraphics[width=\textwidth]{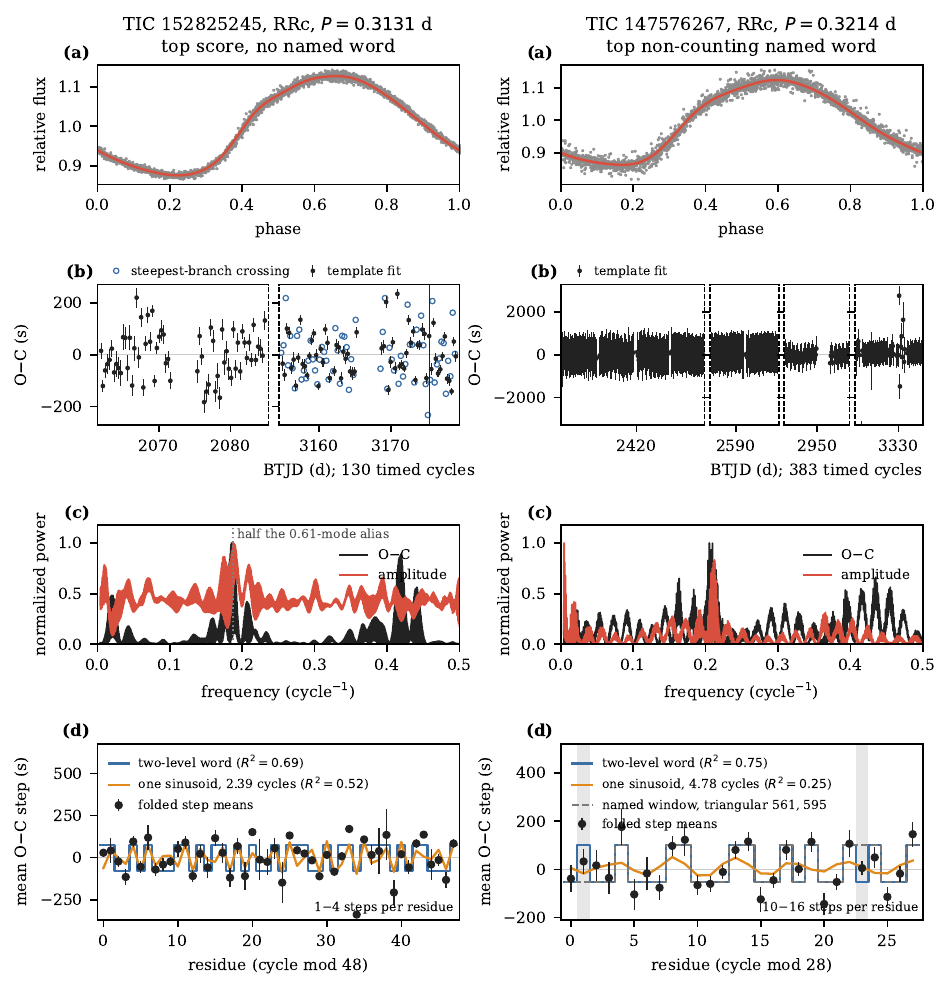}
  \caption{One star from each ordering, shown in bold in its table. Left: TIC~152825245, first by score (Table~\ref{tab:topscore}), with no named word. Right: TIC~147576267, first among the TESS stars of Table~\ref{tab:topseq} once the counting windows are set aside, named as the triangular numbers 561 and 595 in minimal binary. (a) Phase-folded TESS light curve and template. (b) Per-cycle \oc{} from the template fit (filled) and, where stage~2 ran, from the independent steepest-branch crossing time (open); TIC~147576267 was never promoted to stage~2. (c) Lomb--Scargle periodograms of the per-cycle \oc{} (black) and amplitude (red), each normalized; the dotted line marks half the per-cycle alias of the prewhitened 0.61 mode. (d) \oc{} steps averaged per residue of the word length $N_{\rm bit}$, with the decoder's two-level word (blue), the best free sinusoid folded the same way (orange) and, on the right, the named window at its best alignment (grey dashed; shaded bits disagree with the decoded word). $R^2$ is the count-weighted variance of the folded means each model explains. Produced by \texttt{figures/make\_candidates.py}.}
  \label{fig:candidates}
\end{figure*}

Figure~\ref{fig:candidates} shows one star from each ordering. The first is the star at the top of the score ranking. The second is not the first row of Table~\ref{tab:topseq}: six of those ten rows are counting windows, and Section~\ref{sec:topseq} traces that family to the matcher rather than to the stars. We show instead the most significant word outside it that has per-cycle timing. The one above it, ZTF~685111300023197, carries the same triangular window but is block-timed, so it has no per-cycle \oc{} to draw.

\paragraph{TIC 152825245.} TIC~152825245 is a first-overtone RR~Lyrae star (TESS magnitude $T=12.45$, Gaia $G=12.60$) with $P=0.31310$~d, observed by TESS in two sectors three years apart at 45 points per cycle. Its 130 timed cycles have an \oc{} scatter of 83~s against a median formal error of 25~s. Stage~1 flags it on the Ljung--Box and autocorrelation statistics at the floor of their permutation null. Stage~2 cannot remove it. The structure survives cycle boundaries shifted by half a period and a 12-harmonic template ($p=0.002$ for both). The steepest-branch crossing times follow it (Spearman $\rho=0.72$, $p=10^{-11}$; Figure~\ref{fig:candidates}b), each sector is significant on its own, and the block bootstrap puts it above the short-range correlation at the floor of its null. It passes every robustness test and no veto applies, so the star ranks first.

None of those tests looks for a code, and the tests that do look for one find none. The description-length decode reaches only $p=0.41$ and the quantised-step statistic $p=0.53$. The decoder's best word has 48 bits, sampled 2.7 times in 130 cycles, which is too few for the squareness test to run. It is named by no catalog. Decoded separately, the two sectors give unrelated words of 29 and 16 bits. The two-level word explains 69 per cent of the variance of the folded means against 52 per cent for one sinusoid (Figure~\ref{fig:candidates}d), but with only one to four steps per residue that comparison is not evidence for the word: splitting any set of numbers into two levels at the best threshold explains $2/\pi$, or 64 per cent, of their variance when they are Gaussian noise. What the \oc{} does carry is a periodicity. Its periodogram peaks at 0.186~cycles$^{-1}$ (false-alarm probability 0.007), half the per-cycle alias of the 0.1925~d mode that prewhitening removed at 0.25 per cent of the flux (Section~\ref{sec:topscore}). The amplitude periodogram peaks within 0.003~cycles$^{-1}$ of it but is not significant on its own, which is why the common-frequency veto does not fire. Read as the light-travel time of an orbit, the same signal, a 1.67~d sinusoid of 52~s semi-amplitude, would need a mass function of $\sim$55~\Msun{}.

\paragraph{TIC 147576267.} TIC~147576267 is a fainter first-overtone star ($T=13.21$, $G=13.50$) with $P=0.32144$~d, observed in four sectors over 950~d at 46 points per cycle. Its 383 timed cycles have an \oc{} scatter of 375~s, 0.4 times the median formal error of 914~s. That formal error is the fit error inflated by $\sqrt{\chi^2_\nu}$ (Section~\ref{sec:dense}), and here the inflated error exceeds the scatter of the shifts. The steps fold into a 28-bit word that the matched filter names as the triangular numbers 561 and 595 in minimal binary, inverted and rotated by three bits, with 2 of the 28 bits disagreeing. The amplitude $p$-value is $6\times10^{-9}$ and the coincidence $p$-value $1.3\times10^{-3}$, which is the $p$ reported. Each residue of the fold is sampled 10 to 16 times; only TIC~204639317, among the words of Table~\ref{tab:topseq}, is sampled as often, so this is one of the two cases in which the fold-first decoder has the most repetition to integrate.

The case against a message rests on the tests that stage~1 ran, because stage~2 never ran on this star.
\begin{enumerate}
\item The word is not square. Its squareness, and that of the catalog window, is 0.00 against the gate of 0.25. The decode statistic is at the floor of its null ($p=1/2001$), but the decode route to stage~2 also needs a square word, and so does the catalog route, which the window's $p$ of $1.3\times10^{-3}$ would otherwise have cleared. No route promotes it, and the Ljung--Box statistic, at $p=0.004$, misses the $10^{-3}$ gate. The star is QUIET, with a score of 0.17 (quiet base 0.06, a named-word bonus of 0.10, and a small evidence term).
\item The fit is what noise gives. The two-level word explains 75 per cent of the variance of the folded means and the named window 72 per cent, against 25 per cent for the best sinusoid, at 4.78 cycles (Figure~\ref{fig:candidates}d). The two-level figure is close to the $2/\pi$ that a best split of 28 Gaussian means would reach, and the window matches the split rather than adding to it. The \oc{} periodogram has no significant peak (false-alarm probability 0.94). The amplitude periodogram, whose strongest peak outside the secular term is at 0.213~cycles$^{-1}$ (false-alarm probability 0.016), and the step sinusoid both sit where the \oc{} periodogram peaks, 0.21~cycles$^{-1}$, so what little coherent structure the timing has is the kind the common-frequency veto was built for.
\item The word is not reproduced. Decoded separately, the two halves of the baseline give words of 43 and 49 bits, and neither is named.
\item The match reflects the catalog, not the star. The same window is named in ten light curves across the store, two of them in Table~\ref{tab:topseq}, and at $1.3\times10^{-3}$ the star is more than three orders of magnitude from the across-star threshold of Section~\ref{sec:decode}.
\end{enumerate}

The two stars fail as candidates in opposite ways. TIC~152825245 passes every test that can be applied to it and is explained anyway, by the alias of a mode that the tests were not built to single out. TIC~147576267 carries the one thing that a beacon is designed to show, a name from a shared catalog, repeated some fourteen times over the baseline, and every other test shows the name is a coincidence. Taken together they mark out what a real detection would need. It would need a named word that the decoder also finds on its own, squareness well above that of a free sinusoid, a word that repeats in each half of the baseline, and no amplitude modulation at the frequency of the steps. It would need all four in the same star. No star in the store comes close.

\subsection{Transmitter scale at the threshold}
\label{sec:needed}

Since no star is a detection, the useful number is the sensitivity of Table~\ref{tab:sensitivity}, expressed as the smallest transmitter that we would have detected. For the reference RRab star of Table~\ref{tab:physics}, the 50 per cent thresholds of $\dtp=2.4\times10^{-3}$ (primes) and $4.0\times10^{-3}$ (a random 8-bit word) are advances of 118 and 197~s per set bit, or $7\times10^{39}$ and $1.2\times10^{40}$~erg per bit through the resonant annulus at unit gain (Equation~\ref{eq:ebit}). At a duty cycle of one half that is $\pbeam/L=0.39$--$0.66$, which needs 39--66 per cent of a Dyson sphere at $\eta=1$ and four to seven spheres at $\eta=0.1$ (Equation~\ref{eq:ceiling}). Against the $G=1$ kick ceiling, $\dtp=3\times10^{-3}$, the thresholds sit at 0.8 and 1.3: the TESS RR~Lyrae arm can detect, to within a factor of two, the largest code that a self-powered RRab transmitter can send, and only if that transmitter encloses most of its star at close to perfect efficiency. A beacon on the RR~Lyrae stars TESS has timed, built as Section~\ref{sec:physics} describes and signalling at the scale \citet{Learned2008} proposed, would have been found in about half of the stars that carried it.

\section{Limits of a Null Result}
\label{sec:compression}

Every discriminant this search uses is a form of redundancy. The fold integrates repetition, the matched filter looks for sequences that both parties can name, and the squareness statistic looks for a signal with fewer distinct values than a natural process produces. A beacon has to carry that redundancy on purpose: it is meant to be recognized by a receiver that does not know it is there. A message sent to a receiver that already knows the code has no such need. Its sender is better served by removing the redundancy.

Information theory says where that leads. A perfectly compressed message has maximal entropy per symbol, so its bits are statistically indistinguishable from fair coin flips. It does not repeat, it resembles no catalog, and its autocorrelation is that of white noise. \citet{Lachmann2004} showed that a signal sent at the maximum efficiency the physics allows has the statistics of blackbody radiation to anyone without the code. \citet{Messerschmitt2015} reached the corresponding conclusion for power-efficient interstellar signal design. Timing modulation is no exception. On a channel with additive Gaussian noise the input that reaches capacity is itself Gaussian, so a transmitter free to choose the size of each advance, not just whether to make one, would produce Gaussian steps. A clock whose steps are Gaussian is a random walk, which is exactly the Eddington--Plakidis behavior of the stars' own period noise (Section~\ref{sec:jitter}). Strictly, the timing channel here is one-sided and peak-limited, and for an amplitude-constrained Gaussian channel the capacity-achieving input is discrete rather than Gaussian \citep{Smith1971}, so an efficient transmitter with a fixed largest kick would use a few levels rather than a continuum; its bit sequence would still be fair-coin and unrepeated, so the fold, the catalogs and the autocorrelation routes would still see nothing. Even the one-sided binary kick of Section~\ref{sec:onesided}, once its bits are compressed, leaves only the two-level step distribution, and at the kick ceiling that is buried in the star's own wander.

Our result should be read with that in mind. We find no beacon: no star carries a recognizable, repeated, two-level timing code at the amplitudes of Table~\ref{tab:sensitivity}. We cannot say that no star carries a message. A civilization using pulsating stars to communicate efficiently, not to advertise, could write timing residuals that pass every test in this paper. Such a star could be among the 6823 SURVIVORs, whose structure we could not explain, or among the stars whose clocks look merely noisy. Nothing in its timing alone would single it out. One caveat narrows this limit. A compressed code at the amplitudes of Table~\ref{tab:sensitivity} is invisible to the pattern tests but not to the variance: it adds white timing scatter of $(\delta t/2)^2$ per cycle, and a 118--197~s kick would raise the 20--40~s per-cycle scatter of a quiet TESS RR~Lyrae star by a factor of about two to five, which a population test of $\sigma_{\rm O-C}$ against class, period and amplitude would flag as an outlier. This pass did not run that test, so an efficient message at the amplitudes that we can detect would have appeared as excess white jitter, which we did not look for; at the amplitudes a partial swarm can afford, inside the star's own wander, the conclusion above stands. A search for artificial structure in noise-like data is a search for designed inefficiency, and its null result bounds only the civilizations that chose to be inefficient in a way we can recognize.

\section{Conclusions}
\label{sec:conclusions}

We have carried out the automated, survey-scale search for phase-coded pulsators that \citet{Hippke2015} recommended, across the Galactic Cepheids, RR~Lyrae and \bcep{} stars that Gaia~DR3 catalogs and TESS, Kepler, ASAS-SN and ZTF can time. Our conclusions are as follows.

\begin{enumerate}
\item The beacon of \citet{Learned2008} is, in the model they propose, a one-sided, cumulative phase code. A limit cycle never undoes a phase advance, so the observable is a staircase in the \oc{} whose first differences are the bits, and a search must decode increments, not levels.
\item A central neutrino beam at unit gain needs ten to a thousand times the luminosity of the star it drives, because only $10^{-4}$ to $10^{-6}$ of the star's mass lies above the base of the driving zone. A resonant $\bar\nu_e$ beam at the Glashow resonance grazing that layer needs $8\times10^{-3}$ of the luminosity for a \bcep{} star, $2\times10^{-2}$ for a Cepheid and $0.16$ for an RR~Lyrae star. That is the only self-consistent transmitter we have found.
\item RR~Lyrae stars are both the cheapest to drive and the only class to which the present data are sensitive. Injections into real TESS light curves recover half of the codes at $\dtp=2.4$--$4.0\times10^{-3}$, within a factor of two of the largest code a storage-free RR~Lyrae transmitter can send from the whole of its star's light at unit efficiency and unit gain. The Cepheid and \bcep{} arms reach half completeness at no amplitude that is still a phase code. On the ground-based surveys, which hold 82 per cent of the light curves, a one-bit-per-cycle code is invisible; a code held for 32 cycles per bit is flagged at $\dtp\approx(0.6$--$2.6)\times10^{-3}$ and never read.
\item We timed 97{,}725 light curves of 77{,}596 stars and vetted 38{,}230 timing anomalies. The 6823 survivors are almost all ground-based jitter candidates, RRab stars in the proportion the Blazhko effect predicts; the eight with per-cycle timing are aliases of prewhitened non-radial modes, mode beating in \bcep{} stars, one unexplained RRab star, and TIC~235416406 and TIC~347799445, whose unresolved beats survive because the \oc{} and the amplitude do not peak together.
\item No decoded word survives a false-discovery correction across stars. The most significant words are the per-cycle alias of a non-radial mode matched against a near-periodic stretch of the counting catalog, repeated in 91 unrelated stars, and bit-flipped and reversed catalogs name stars as often as the real one. We find no beacon.
\item A perfectly compressed message is indistinguishable from noise, so this null result constrains beacons, not communication. Stars carrying efficiently encoded signals may pass every test applied here, and we cannot single them out.
\end{enumerate}

The next pass can be sharper. Still to do are a coincidence null for named words drawn from sinusoidal words at the observed alias frequency rather than from uniform arrangements, a catalog prior that weights the start of each short sequence more heavily, so that a genuine 28-bit primes beacon could pass the across-star correction, a population test of $\sigma_{\rm O-C}$ against class, period and amplitude (Section~\ref{sec:compression}), a harmonic-sideband test (a phase modulation's sidebands grow linearly with harmonic order, which no additive second mode reproduces), and a null that accounts for red noise in the Cepheids, whose clocks are at present too noisy for any code to be detected. The codes still out of reach are those of partial swarms: a transmitter with a twentieth of the sky at ten per cent efficiency sends kicks of $\dtp=1.5\times10^{-5}$, 0.7~s against a per-cycle scatter of 20--40~s, and a 28-bit word read through the fold\footnote{It needs $r\approx(3\times30/0.7)^2\approx1.5\times10^{4}$ repeats for a $3\sigma$ residue mean.} then needs some $4\times10^{5}$ timed cycles, six centuries of continuous per-cycle timing of a 0.57~d star whose clock is assumed white throughout. TESS full-frame-image timing for the 8218 RR~Lyrae stars that have no light-curve product and Gaia~DR4 epoch photometry stitched to the ASAS-SN and ZTF baselines are each an order-of-magnitude step in timed cycles toward targets that recede as the assumed swarm shrinks.

We did not find a beacon, but we now know what one would have to be: a one-sided staircase in the \oc{}, written by a resonant beam grazing the base of the driving zone and cheapest on the RR~Lyrae stars. The RR~Lyrae stars that TESS has timed were therefore the right place to search first, and none shows a beacon at the largest amplitude that its own light could power. That null result turns an open proposal into a bounded one, and it shows how much smaller a code the next search must be able to detect.

\section*{Acknowledgements}
The manuscript source, the bibliography, and the equilibrium-solver and figure scripts are available online.\footnote{\url{https://github.com/guillochon/starseti}} The author takes full responsibility for the contents of this paper. Large language models (Claude Fable 5.1, Grok 4.6, and GPT5.6 Sol) were used, under author supervision, to draft and edit the text, to derive and check analytic results, to write and refine the analysis and figure-generation scripts released with this paper, and to refine the presentation of the figures. The figures themselves are produced by those scripts from the solver outputs; no generative model created or altered the plotted data. Every displayed equation has been re-derived independently, every entry in the bibliography was validated against the NASA Astrophysics Data System, and all references and AI edits were hand-checked by the author. We aimed to be compliant with the AAS guidelines on generative AI in manuscript preparation.\footnote{\url{https://journals.aas.org/author-llm-guidelines}} Those guidelines do not treat such tools as authors; responsibility for accuracy remains with the listed author. We thank J. Xavier Prochaska for providing access to AI models that were used to conduct independent AI reviews.

This paper includes data collected by the TESS mission, which are publicly available from the Mikulski Archive for Space Telescopes (MAST). Funding for the TESS mission is provided by the NASA Explorer Program. This work has made use of data from the European Space Agency (ESA) mission Gaia, processed by the Gaia Data Processing and Analysis Consortium (DPAC), of the ASAS-SN Sky Patrol, and of the Zwicky Transient Facility light-curve service at IRSA. ZTF is supported by the National Science Foundation and a collaboration of partner institutions.

\emph{Facilities:} TESS, Gaia, ASAS-SN, PO:1.2m (ZTF), MAST, IRSA.

\emph{Software:} \texttt{astropy} \citep{Astropy2022}, \texttt{lightkurve} \citep{Lightkurve2018}, \texttt{numpy} \citep{Harris2020}, \texttt{scipy} \citep{Virtanen2020}, \texttt{numba} \citep{Lam2015numba}, \texttt{matplotlib} \citep{Hunter2007}.

\section*{Declaration of competing interest}
The author declares no competing financial interests or personal relationships that could have appeared to influence the work reported in this paper.

\section*{CRediT authorship contribution statement}
\textbf{James Guillochon:} Conceptualization, Methodology, Software, Formal analysis, Investigation, Data curation, Visualization, Writing -- original draft, Writing -- review \& editing.

\bibliographystyle{elsarticle-num-names}
\bibliography{references,references-seti}

\end{document}